\documentclass[twocolumn,aps,prb,floats,showpacs,superscriptaddress]{revtex4}
\usepackage{amsmath}

\newcommand{\be}{\begin{equation}}
\newcommand{\ee}{\end{equation}}

\usepackage{epsfig}
\usepackage{graphicx}
\usepackage{bm}
\usepackage{amssymb}%
\usepackage{slashed}

\def\gapp{\lower.35em\hbox{$\stackrel{\textstyle>}{\sim}$}}
\def\lapp{\lower.35em\hbox{$\stackrel{\textstyle<}{\sim}$}}

\newcommand{\Neel}{N\'{e}el}

\begin{document}

\bibliographystyle{apsrev}

\title{Dirac electrons and domain walls: a realization in junctions of ferromagnets and topological insulators}
\author{Yago Ferreiros}
\email{yago.ferreiros@csic.es}
\affiliation{Instituto de Ciencia de Materiales de Madrid, CSIC, Cantoblanco, 28049 Madrid, Spain,}
\author{F. J. Buijnsters}
\email{f.buijnsters@science.ru.nl}
\affiliation{Institute for Molecules and Materials, Radboud University Nijmegen.}
\author{M. I. Katsnelson}
\email{m.katsnelson@science.ru.nl}
\affiliation{Institute for Molecules and Materials, Radboud University Nijmegen.}

\begin{abstract}
We study a system of Dirac electrons with finite density of charge carriers coupled to an external electromagnetic field in two spatial dimensions, with a domain wall (DW) mass term. The interface between a thin-film ferromagnet and a three-dimensional topological insulator provides a condensed-matter realization of this model, when an out-of-plane domain wall magnetization is coupled to the TI surface states. We show how, for films with very weak intrinsic in-plane anisotropies, the torque generated by the edge electronic current flowing along the DW competes with an effective in-plane anisotropy energy, induced by quantum fluctuations of the chiral electrons bound to the wall, in a mission to drive the internal angle of the DW from a Bloch configuration towards a \Neel{} configuration. Both the edge current and the induced anisotropy contribute to stabilize the internal angle, so that for weak intrinsic in-plane anisotropies DW motion is still possible without suffering from an extremely early Walker breakdown.

\end{abstract}

\maketitle

\section{Introduction}

Dirac fermions in two spatial dimensions have been the object of intense study in recent times in the condensed-matter world, especially since the experimental realization of graphene\cite{NGM04,NGP09,K12} and, more recently, the discovery of three-dimensional topological insulators (TI)\cite{FKM07,HK10,XZ11}, which host Dirac fermions as topologically protected surface states. On the other hand, magnetic DWs and their manipulation via applied currents and electromagnetic fields hold a prominent position in the field of spintronics, especially so since the experimental realization of the ``race-track'' technology\cite{PHM08,TMR10}. 

The search of efficient ways of moving a DW at the highest possible velocities has become of capital importance. Manipulation based on the application of external magnetic fields\cite{SW74,OMS99,AAX03,NHO01,TGM02,NTM03,BNK05}, currents\cite{B84,FB85,HB88,GCA00,KKG02,YON04,GBC03}, and more recently magnons\cite{YWW11,KSK12,WGN12,WGZ13,HTH14,MNM14,WAB15} and electric fields\cite{OCM00,SBF11} has been proposed and, with the exception of magnonic manipulation, experimentally realized. There is an upper limit on the DW velocity due to the phenomenon known as Walker breakdown (WB)\cite{SW74}. Above a threshold applied current or magnetic field, the internal structure of the DW, as described by its internal angle, becomes unstable. The net velocity of the DW is limited by this effect. As a consequence, the search of mechanisms that can stabilize the internal angle of the DW has become an important task\cite{GCC05,B13}. In this regard, stabilization mediated by Dzyaloshinskii-Moriya (DM) interactions\cite{TRJ12,SUS13} and Rashba fields\cite{ITH11} has been explored and experimentally realized.

The appearance of three-dimensional TIs has focused the interest on what we could call Dirac-fermion-mediated ferromagnetism\cite{CYO12} and spintronics. The aspect of TIs that makes them highly valuable for spintronic applications is that the spin orientation of the surface electrons is fixed relative to their propagation direction, so that the effects of spin-orbit coupling are maximal. For this reason, some proposals\cite{BH10,CHS10,PM12,FVM13} have suggested that TIs could possess more efficient spin--orbit-induced torques than other materials previously considered. Indeed, the strength of the spin-transfer torque per unit charge-current density, exerted by the TI surface states on the magnetization of an adjacent ferromagnetic permalloy thin film, has recently been measured to be greater than for any other spin-transfer source measured so far\cite{FUK14,MLR14}.

In this context it seems natural and promising to study TIs coupled to ferromagnets\cite{YTN10,YZN10,GF10,Y11,UTT12,NE12,NE14,TPL15} and specifically to DWs\cite{NN10,TL12,WB12,HP13,FC14,L14,WEN14}.
If we couple a TI layer to an out-of-plane magnetized ferromagnetic thin film containing a domain wall, the DW acts as a mass for the surface electrons. It is a realization of a system of Dirac fermions with a DW mass term.
The theory of this system coupled to electromagnetism was studied in a field-theoretical context in Ref.~\cite{CH85}. These authors showed how the $2n$-dimensional anomaly of the chiral fermions living in the DW is canceled by the anomaly due to the induced $2n+1$ topological mass term. This cancellation in three spacetime dimensions has been explicitly computed for fermions coupled to an abelian gauge field in Ref.~\cite{C94}, where the technical difficulties that appear when trying to obtain the effective action in the presence of a DW mass become apparent.

In this Article, we obtain analytical expressions for the effective action for an external electromagnetic field of a system of Dirac fermions in two spatial dimensions, at a finite density of charge carriers and with a DW mass term. We look at a physical realization in junctions of three-dimensional TIs and ferromagnets with out-of-plane easy-axis anisotropy that host an out-of-plane DW. We show how the surface electrons of the TI induce an effective in-plane anisotropy energy, which stabilizes the DW in a Bloch configuration even in thin ferromagnetic films where in-plane intrinsic and dipolar (shape) anisotropies are relatively weak. Owing to the stabilization of the internal angle, DW motion is possible without suffering from a very early Walker breakdown.

We also show how equilibrium and nonequlibrium edge currents can be generated along the DW by applying, respectively, a gate voltage (doping with electrons/holes) or an electric field. This current exerts a torque on the magnetization that drives the internal angle from the Bloch configuration towards a \Neel{} configuration. We analytically compute the behavior of the DW internal angle as a function of the chemical potential and find it to qualitatively agree with a recent numerical calculation\cite{WEN14}. Furthermore, this edge current contributes to the stabilization of the internal degree of freedom of the DW in a similar way as the DM interaction. It turns out that the corresponding term in the effective action, although it is first order in the magnetization, qualitatively resembles that of an interfacial DM interaction. Here the DM interaction is tuneable, with the interaction strength given by the amount of edge current flowing along the DW.

Our approach has the advantage that, since the calculation is done at the microscopic level, the origin of the physics underlying the phenomenology can be traced and well understood.

\section{Effective action}
\label{sec effective act electromagnetic}

In this Section, we present a the calculation of the effective action for the electromagnetic field coupled to a system of Dirac electrons with a DW mass. Readers interested primarily in the physics and phenomenology of ferromagnet--TI junctions may skip to Sec.~\ref{section thin film magnet/TI junctions}.

Let us start from the action of $3d$ Dirac fermions (in the following $nd$ will be used for $n$-dimensional) coupled to an external electromagnetic field:
\begin{equation}
\mathcal{S}=\int d^3x\,\bar{\Psi}\big(i\,l^\mu_\nu\gamma^\nu(\partial_\mu-ie A_\mu)-m-\gamma^0\mu\big)\Psi
\label{eq class act electrodynamics}
\end{equation}
with the DW mass
\begin{equation}
m=m_0\sigma\tanh(x_1/\delta)
\end{equation}
where $x_\mu=(x_0,x_1,x_2)$, $x_0$ is time, $\sigma=\pm1$ and $\delta$ are the topological charge and the width of the DW respectively, we define $\bar{\Psi}=\Psi\gamma^0$, $\mu$ is the chemical potential, $l^\mu_\nu$ is given by
\begin{equation}
l^\mu_\nu=\begin{pmatrix}
1&0&0\\0&v&0\\0&0&v\end{pmatrix}
\end{equation}
and the gamma matrices satisfy the anticommutation relations $\{\gamma^\mu,\gamma^\nu\}=2\eta^{\mu\nu}$. We work with a metric with signature $(+,-,-,-)$ and with $\hbar=c=1$. We consider the general case where the velocity $v$ is not necessarily equal to the speed of light $c$.

We obtain the effective action for $A_\mu$ up to second order in the fields. The fermionic spectrum consists of a chiral massless state bound to the DW plus massive extended states (see Appendix \ref{sec spectrum}). If the DW is wide enough, massive bound states also appear. The total number of bound states is given by the largest integer less than $\lambda+1$, with the parameter $\lambda$ given by
\begin{equation}
\lambda=\frac{m_0\delta}{v}
\label{eq lambda}
\end{equation}
(see Appendix~\ref{sec spectrum}). Let us consider the case where we have a DW that is so steep that the only bound state is the chiral state ($\lambda\leq1$), so that we can do an enlightening separation. We can consider the system as described by two theories, one 2$d$ edge theory, describing the chiral electrons localized near the DW center, and one 3$d$ bulk theory describing the massive extended electrons. Each of these theories, considered in isolation, is anomalous. The 2$d$ chiral edge theory, on the one hand, is well known to be chiral anomalous\cite{JR85}, while the anomaly in the bulk theory is a consequence of the generation [via a Chern-Simons (CS) term] of a topological mass of opposite signs on either side of the DW. However, the anomalies cancel via the Callan--Harvey mechanism \cite{CH85}, so that the complete theory is anomaly free.

The perturbative calculation of the effective action would in principle require computing the fermionic propagator from the exact fermionic spectrum, and then performing the integration of the fermionic degrees of freedom in the path integral. This was done to second order in $A_\mu$ and for $\mu=0$ in Ref.~\cite{C94} as an explicit verification of the anomaly cancellation, but these authors considered only the case $m_0\rightarrow\infty$ and focused exclusively on those terms that contain either two-dimensional or three-dimensional antisymmetric tensors, which are the terms relevant for the anomalies. The complete analytic computation for finite $m_0$ and $\mu$ remains a formidable task.

To make the calculation tractable, we must introduce a number of approximations. First, we consider the adiabatic limit, assuming a constant mass in the calculations and restoring the $x_1$ dependence at the end. This approximation captures the bulk contribution (extended states) and is reliable as long as the energy associated with the typical length of the inhomogeneities in the mass is much smaller than the energy of the extended states ($\sim m_0$). This means that the approximation is asymptotically exact, but near the DW center it translates into the condition $v/\delta \ll m_0$ which is never fulfilled if we consider the case $\lambda\leq1$. As a consequence, nonadiabatic corrections will appear near the DW. Furthermore, even if the condition is fulfilled, this approach cannot describe the contribution of the bound states. Hence, as a second approximation, we add the contribution of the chiral state, assuming $\lambda\leq1$ to avoid further computations for the contribution of the massive bound states. As a third and final approximation, we compute non-adiabatic corrections to the CS term of the bulk contribution. To obtain these corrections, we impose gauge invariance and the cancellation of the anomalies in the two theories (bulk--edge correspondence).

\subsection{Edge theory}

Let us first compute the 2$d$ edge effective action. From the action of eq. (\ref{eq class act electrodynamics}) and the fermionic spectrum obtained in Appendix \ref{sec spectrum}, the classical action for the chiral mode can be written as:
\begin{equation}
\mathcal{S}_{R,L}=\int d^3x\,\rho_\lambda^2(x_1)\,{\psi^{{(0)}^*}_{R,L}}(x_0,x_2)\Big(i\partial_0+eA_0+
\nonumber
\end{equation}
\begin{equation}
+\sigma v\big(i\partial_{2}+eA_2\big)-\mu\Big)\psi^{(0)}_{R,L}(x_0,x_2)
\label{Chiral act}
\end{equation}
where (see eq. (\ref{eq. chiral mode})):
$$
\rho_\lambda(x_1)=B_0(\lambda)\cosh^{\lambda+1}(x_1/\delta)\times
$$
\begin{equation}
\times{_2}F_1\Big[\frac{1}{2},\lambda+\frac{1}{2},\frac{1}{2};-\sinh^2(x_1/\delta)\Big]
\label{eq rho}
\end{equation}
with $B_0(\lambda)$ defined in eq. (\ref{eq B0}) and ${_2}F_1$ the hypergeometric function. When $\sigma=-1$ we have $\mathcal{S}_{R}$ and when $\sigma=1$ we have $\mathcal{S}_{L}$.

Before proceeding, we partially fix the gauge. Let us set
\begin{equation}
A_\mu\rightarrow\theta^{(\mu)}(x_1)A_\mu(x_0,x_2)
\label{eq gauge cond}
\end{equation}
where $\theta^{(\mu)}(x_1)$ is a given function of $x_1$. The remaining gauge freedom is given by
\begin{equation}
A_\mu\rightarrow A_\mu+\partial_\mu\omega
\end{equation}
with $\omega\neq\omega(x_1)$.

\subsubsection{Equilibrium edge current}

As a consequence of having a finite chemical potential, when integrating out the chiral fermion in eq. (\ref{Chiral act}) the tadpole terms do not vanish. This gives linear terms in the effective action and an associated equilibrium (external electromagnetic fields set to zero) chiral edge current density along the DW. The linear terms can be computed from the tadpole-like term (in imaginary time):
$$
\Gamma_{eq}=-ie\int d^3x\,\rho_\lambda^2(x_1)\,(iA_0+\sigma v A_2)\times
$$
\begin{equation}
\times\int \frac{dq_0dq_2}{4\pi^2}\frac{1}{q_0+\sigma ivq_2-i\mu}
\end{equation}
By multiplying and dividing by $q_0-\sigma ivq_2+i\mu$ the integral in momenta can be rewritten as an even plus an odd part in $q_0$. The integration of the odd part vanishes, while the integration over both momenta of the even part gives (going back to real time):
\begin{equation}
\Gamma_{eq}=-\frac{e\mu}{2\pi v}\int d^3x\,\rho_\lambda^2(x_1)\, (A_0+\sigma vA_2)
\label{action chiral eq}
\end{equation}
so that we can define the equilibrium edge current density as:
\begin{equation}
j^a_{eq}=-\frac{e\mu}{2\pi v}\rho_\lambda^2(x_1)\,(1,\sigma v)
\label{eq eq current}
\end{equation}
Here and from now on the Greek letters refer to dimensions $x_0$, $x_2$: $a=0,2$.

\subsubsection{Chiral anomaly}

The action of eq. (\ref{Chiral act}) is analogous to that of the 2$d$ chiral Schwinger model, which is well known to be chiral anomalous\cite{JR85}, so that the gauge symmetry at the quantum level is broken. Integrating out the chiral fermionic degrees of freedom in eq. (\ref{Chiral act}) up to second order in the electromagnetic field we get (using dimensional regularization)\cite{JR85}:
\begin{widetext}
$$
\Gamma_{anomaly}=\frac{1}{4\pi v}\int dx_1dx'_1\rho_\lambda^2(x_1)\rho_\lambda^2(x'_1)\times
$$
\begin{equation}
\int dx_0dx_2\,A_{a}(x_0,x_1,x_2)\bigg(\eta^{ab}-\frac{\partial^{a}\partial^{b}}{\partial^{2}}-\frac{\sigma}{2\partial^{2}}(\epsilon^{cb}\partial^{a}\partial_{c}-\epsilon^{ad}\partial_{d}\partial^{b})\bigg)A_{b}(x_0,x'_1,x_2)
\label{eq anomaly}
\end{equation}
\end{widetext}
where $\partial_a=(\partial_0,v\partial_2)$, $A_a=(A_0,vA_2)$ and $\partial^2=\eta^{ab}\partial_a\partial_b$. Note that chemical potential does not play a role here, and there are two reasons for this to happen. First for massless (non chiral) 2$d$ fermions the theory at finite charge density is indistinguishable to that at zero density. Second, the chiral anomaly is well known to be insensitive to chemical potential and temperature\cite{HM83,DK87}. However, as we obtained in eq. (\ref{action chiral eq}) finite $\mu$ plays a role at first order in $A_\mu$ for chiral 2$d$ fermions. The effect is equivalent to applying a chiral chemical potential to non-chiral and massless fermions, which activates the chiral magnetic effect in 2$d$ generating an equilibrium current density analogous to that of eq. (\ref{eq eq current}).

Finally, from eq. (\ref{eq anomaly}) we can obtain the nonequilibrium edge current density:
\begin{equation}
j^a_{ne}=-\sigma\frac{e^2}{2\pi }\rho_\lambda^2(x_1)\int dx'_1\frac{\rho_\lambda^2(x'_1)E_2(x_0,x'_1,x_2)}{\partial_0+\sigma v\partial_2}(1,\sigma v)
\label{eq noneq current}
\end{equation}
with $E_2=\partial_0A_2-\partial_2A_0$ the electric field in the $x_2$ direction.

\subsection{Bulk theory}

To obtain the bulk contribution we will assume a constant mass for the fermions, restoring the $x_1$ dependence at the end of the calculations in what basically is an adiabatic approximation, as we mentioned before. This way we have 3$d$ Dirac fermions with a mass term that breaks time reversal symmetry. This system is well known to give a topological response under an external electromagnetic field in the form of a CS term\cite{R84}. 

To proceed we can always split the effective action into vacuum ($\mu=0$) and matter ($\mu\neq0$) contributions, so that the matter contribution is zero at zero density:
\begin{equation}
\Gamma_{bulk}=\Gamma_0+\Gamma_{matt}
\end{equation}
with:
\begin{equation}
\Gamma_{0,matt}=\frac{1}{2}\int d^3x\, A_\mu\,\Pi_{0,matt}^{\mu\nu}\,A_\nu
\end{equation}
where $A_\mu=(A_0,vA_1,vA_2)$ and where $\Pi^{\mu\nu}$ is the polarization function. Doing this separation all ultraviolet divergences appear in the vacuum part, while the matter part remains finite.

\subsubsection{Vacuum contribution}

To be consistent with the calculations done for the edge theory, we will use dimensional regularization to treat the ultraviolet divergences. The computation of the 1-loop polarization function for $\mu=0$ is straight forward. Separating it into it's even and odd parts:
\begin{equation}
\Pi_0^{\mu\nu}=\Pi_{0,e}^{\mu\nu}+\Pi_{0,o}^{\mu\nu}
\end{equation}
and doing the computation we obtain:
\begin{equation}
\Pi_{0,e}^{\mu\nu}=\frac{e^2 |m|}{12\pi v^2}\Big(\frac{\partial^2}{m^2}+\mathcal{O}\big(\frac{\partial^4}{m^4}\big)\Big)\Big(\eta^{\mu\nu}-\frac{\partial^\mu\partial^\nu}{\partial^2}\Big)
\label{eq vacuum even}
\end{equation}
\begin{equation}
\Pi_{0,o}^{\mu\nu}=-\frac{e^2\, sign(m)}{4\pi v^2}\epsilon^{\mu\rho\nu}\partial_\rho\Big(1+\mathcal{O}\big(\frac{\partial^2}{m^2}\big)\Big) 
\label{eq CS}
\end{equation}
with $m=m_0\sigma\tanh(x_1/\delta)$. Here $\partial_\mu=(\partial_0,v\partial_1,v\partial_2)$ and $\partial^2=\eta^{\mu\nu}\partial_\mu\partial_\nu$. Note that we presented the results as the first terms in a derivative expansion, which will be useful later on when we treat the physical system. This expansion is justified in the low energy regime $p^2<<m^2$, breaking down when $m^2\lesssim p^2$. If $p^2<<m_0^2$ it turns out that this breakdown occurs in the region near the DW where the adiabatic limit is no longer reliable. Hence the validity of the derivative expansion coincides with the validity of the adiabatic approximation. 

Now let us look at the gauge variation of the full theory (edge plus bulk). The gauge variation of the edge theory (eq. (\ref{eq anomaly})) is:
\begin{equation}
\delta\Gamma=\frac{\sigma e^2}{4\pi}\int d^3x\, \omega \rho_\lambda^2(x_1) E_2
\label{eq gauge edge theory}
\end{equation}
which should be canceled by the gauge variation of the CS term (eq. (\ref{eq CS})) so that the theory is gauge invariant. This is not the case if $E_2$ is a function of $x_1$. The reason is that while the nature of the edge theory is totally non-adiabatic, the bulk theory has been computed in the adiabatic approximation. To cure this issue, we write the CS term as:
\begin{equation}
\Pi_{0,o}^{\mu\nu}=-\frac{e^2\sigma F_\lambda(x_1)}{4\pi v^2}\epsilon^{\mu\rho\nu}\partial_\rho
\end{equation}
so that the non-adiabatic term $F_\lambda(x_1)$ is fixed by imposing the anomaly cancellation. This is done in Appendix \ref{sec non adiab CS}, where the explicit form of $F_\lambda(x_1)$ can be found.

\subsubsection{Matter contribution}

The computation of the matter contribution is more involved. We again do the separation:
\begin{equation}
\Pi_{matt}^{\mu\nu}=\Pi_{matt,e}^{\mu\nu}+\Pi_{matt,o}^{\mu\nu}
\end{equation}
The even part is computed in Appendix \ref{sec matter contr bulk}, where adding both vacuum and matter contributions we get:  
$$
\Pi_{e}^{00}=-\theta(m^2-\mu^2)\frac{e^2|\bm\partial|^2}{12\pi|m|}+
$$
\begin{equation}
+\theta(\mu^2-m^2)\frac{e^2(|\mu|-|m|)}{2\pi v^2}
\end{equation}
\begin{equation}
\Pi_{e}^{0i}=\Pi_{e}^{i0}=\theta(m^2-\mu^2)\frac{e^2\partial^0\partial^i}{12\pi v|m|}
\end{equation}\begin{equation}
\Pi_{e}^{ij}=-\theta(m^2-\mu^2)\frac{e^2\partial^2}{12\pi v^2|m|}\Big(\delta^{ij}+\frac{v^2\partial^i\partial^j}{\partial^2}\Big)
\label{eq bulk even}
\end{equation}
Some approximations have been done to arrive to these expressions. First, we are in both the adiabatic and the low energy limits. Second, the result is obtained by adding the polarization functions computed in both the static ($p_0\rightarrow0$) and the long wavelength ($\mathbf{p}\rightarrow0$) limits. As explained in Appendix \ref{sec matter contr bulk}, this means that within our approximations non-local terms which are constant in the limit $p_0\rightarrow0$ and zero in the limit $\mathbf{p}\rightarrow0$ are being approximated by the constant term in $\Pi_{e}^{00}$. Third, we assume that in the static limit the spatial momentum $|\mathbf{p}|$ is smaller than the Fermi momentum $p_F$. This is generally true at low energies, except when the Fermi energy $|\mu|$ is above but very close to the value of $|m_0|$, so that $p_F$ is very small. In this situation some corrections which are highly non-local would contribute. And fourth, we neglect dynamical contributions which are non-local (inverse powers in the spatial momentum), which we believe will not have an appreciable effect in the description of the physical system, as we acknowledge in Appendix \ref{app in plane second order}.

To complete the bulk part of the effective action let us turn our attention to the odd part of the polarization function, this is to the matter correction to the CS term. The calculation is straight forward and we obtain (vacuum plus matter contributions):
$$
\Pi_{o}^{\mu\nu}=-\frac{e^2}{4\pi v^2}\Big(\sigma F_\lambda(x_1)\,\theta(m^2-\mu^2)+
$$
\begin{equation}
+\frac{m}{|\mu|}\theta(\mu^2-m^2)\Big)\epsilon^{\mu\rho\nu}\partial_\rho
\label{eq bulk odd}
\end{equation}
Note that all the non-adiabaticity is encoded in $F_\lambda(x_1)$, whereas no non-adiabatic corrections have been computed for $\mu^2>m^2$. The reason is that corrections in this last case can not be computed as we did for $\mu^2<m^2$, as the anomaly cancellation can not be invoked. The gauge non-invariance of the CS term for $\mu^2>m^2$ can not be canceled by the chiral anomaly, which is insensitive to the chemical potential (and temperature $T$)\cite{HM83,DK87}. Furthermore, for finite $\mu$ and/or $T$ there appear infinitely many terms in the perturbative series (in $A_\mu$) that break gauge invariance in the presence of a boundary (or a DW mass as in our case). These terms are parity breaking (the CS is the lowest order of this terms), and in the presence of a boundary are gauge invariant for zero $\mu$ and $T$ (the only exception to this is the CS term) and gauge non-invariant for finite $\mu$ and/or $T$ (see\cite{BDF00} for a computation of the next order parity breaking term for finite $T$). Therefore gauge invariance can only be restored at the non-perturbative level when all the terms in the perturbative series are summed up, in a similar way as occurs with large gauge invariance in the theory with no boundaries\cite{DLL97,DGS97,BDF01}.

\section{Junction of a ferromagnet and a topological insulator}
\label{section thin film magnet/TI junctions}

Let us now look at a condensed-matter realization of the theory of Sec \ref{sec effective act electromagnetic}. We will take a thin film of a ferromagnet and place it on top of a three-dimensional TI. The action for the ferromagnet is (we restore $\hbar$ for the rest of the main text):
\begin{equation}
S_{FM}=d\int dtdxdy\Big(\frac{M_s}{\gamma}\dot{\phi}(\cos\theta-1)-H_{FM}\Big)
\label{eq FMaction}
\end{equation}
which is the sum of the Berry phase term (``kinetic energy of spin precession") plus the Hamiltonian:
\begin{equation}
H_{FM}=\frac{1}{2}\left[A\left(\left|\frac{\partial\mathbf{m}}{\partial x}\right|^2+\left|\frac{\partial\mathbf{m}}{\partial y}\right|^2\right)-Km_z^2+K_\perp m_x^2\right]
\label{FMham}
\end{equation}
(see for example\cite{TKS08,STK11}). Here $d$ is the film thickness, $\gamma=\mu_Bg_e/\hbar$ the gyromagnetic ratio ($g_e=2$), $M_s$ the saturation magnetization, $A$ the exchange constant (exchange energy per unit length) and $K$ and $K_\perp$ the easy axis and hard axis anisotropy constants (anisotropy energy per unit volume). The magnetization unit vector is $\mathbf{m}=(\sin\theta\cos\phi,\sin\theta\sin\phi,\cos\theta)$. It relates to the total magnetization as $\mathbf{M}=\hbar\gamma\,\mathbf{m}/a^3=M_s\,\mathbf{m}$ (where $a$ is the lattice constant) and couples to the spin of the surface electrons of the TI insulator via an exchange interaction.

The action for the TI surface electrons, including the coupling to the magnetization, takes the standard form:
\begin{equation}
\mathcal{S}_{TI}=\int dtdxdy\,\Psi^{\dagger}(i\hbar\partial_0-H_{TI}-\mu)\Psi
\label{TIact}
\end{equation}
with
$$
H_{TI}= v_{F} \,\hat{\mathbf{z}}\cdot(i(\hbar\nabla-ie\mathbf{A})\times\bm\sigma) \pm
$$
\begin{equation}
\pm\Delta_{xy}\mathbf{m}_{xy}\cdot\bm\sigma_{xy}\pm\Delta_z m_z\sigma_z-eA_0, \label{TIham}
\end{equation}
where the surface of the TI is taken to be in the $z=0$ plane, and where we have defined the density to be zero at half-filling. $\Delta_{xy}$ and $\Delta_z$ are the in-plane and out-of-plane exchange couplings respectively (both are definite positive); $v_F$ is the Fermi velocity of the electrons; $\mathbf{m}=(m_x,m_y,m_z)=(\mathbf{m}_{xy},m_z)$ and $\bm\sigma=(\sigma_x,\sigma_y,\sigma_z)=(\bm\sigma_{xy},\sigma_z)$; $A_\mu=(A_0,\mathbf{A})$ and the signs $+$ and $-$ correspond to antiferromagnetic and ferromagnetic exchange couplings, respectively. $S_{TI}$ can be written as eq. (\ref{eq class act electrodynamics}) (with the representation for the Gamma matrices given by (\ref{eq gamma})) setting $x^\mu=(t,x,y)$, $v=v_F$, $m=\pm\Delta_z m_z$ and doing the substitution: $eA_\mu\,\rightarrow\,e\,a_\mu=(eA_0\, ,eA_x\pm\Delta_{xy}m_y/v_F\, ,eA_y\mp\Delta_{xy}m_x/v_F)$. 

The Hamiltonian (\ref{FMham}) supports a non trivial minimum energy configuration in the form of the well known Bloch DW:
\begin{equation}
\theta_0=2\arctan(e^{\sigma x/\delta}),\quad\phi_0=\pm\pi/2
\end{equation}
where $\delta=\sqrt{A/K}$ and $\sigma=\pm1$ are the width and the topological charge of the DW respectively. This is valid if $K_\perp\neq0$, while if $K_\perp=0$ the vacuum would be degenerate and $\phi_0$ could take any value. This configuration gives the equilibrium magnetization:
\begin{equation}
\mathbf{m}^{(0)}=\big(sech(x/\delta)\cos\phi_0,sech(x/\delta)\sin\phi_0,\sigma\tanh(x/\delta)\big)
\label{eq equilibrium mag}
\end{equation}
We will work with fluctuations around the equilibrium configuration: $\theta(t,x)=\theta_0(x)+\tilde{\theta}(t,x)$ and $\phi(t,x)=\phi_0+\tilde{\phi}(t,x)$, giving the total magnetization:
\begin{equation}
\mathbf{m}=(\sin\theta\cos\phi,\sin\theta\sin\phi,\sigma\tanh(x/\delta)+\tilde{m}_z)
\label{eq total mag}
\end{equation}
with $\tilde{m}_z=\cos\theta-\cos\theta_0$. Note that we have imposed the magnetization to be homogeneous in the $y$ direction.

The gauge fixing condition given by eq. (\ref{eq gauge cond}) has now to be fulfilled by the effective vector potential $a_\mu$. We fix the gauge:
\begin{equation}
ea_0=-eE_xx-eE_yy
\label{eq effective gauge 1}
\end{equation}
\begin{equation}
ea_x=\pm \Delta_{xy}m_y(t,x)/v_F
\label{eq effective gauge 2}
\end{equation}
\begin{equation}
ea_y=\mp\Delta_{xy}m_x(t,x)/v_F
\label{eq effective gauge 3}
\end{equation}
with:
\begin{equation}
m_x(t,x)=\sin(\theta_0(x)+\tilde{\theta}(t))\cos\phi(t)
\end{equation}
\begin{equation}
m_y(t,x)=\sin(\theta_0(x)+\tilde{\theta}(t))\sin\phi(t)
\end{equation}
where we have chosen an electrostatic configuration of the electromagnetic field ($E_{x,y}$ are constants). Since the $x$ dependence of $\mathbf{m}_{xy}$ has been fixed, the fluctuations can only be functions of $t$. To obtain an effective action for $x$ dependent fluctuations we have to relay on the adiabatic approximation, restoring the $x$ dependence at the end. For the adiabatic approximation to be valid the wavelength of the spin waves $l_{sw}$ in the $x$ direction has to be much bigger than the typical wavelength of the surface electrons $l_{el}=\hbar v_F/\Delta_z$. Assuming the parameter $\lambda=\Delta_z\delta/(\hbar v_F)$ introduced in eq. (\ref{eq lambda}) that defines the number of existent bound surface states (see Appendix \ref{sec spectrum}) to be $\lambda\sim1$, we have $l_{sw}>>\delta$.


This way, we have "almost" a completely analogous theory for the fermionic sector to that in the previous section, but now for the effective electromagnetic field $a_\mu$. We say "almost" because in addition we have an extra field $\tilde{m}_z$ with which we have to deal. Before proceeding, let us fix the values of the parameters. We set (the MI parameters are obtained from\cite{MLA09}): $d=3nm$, $M_s=3\times 10^5A/m$, $A=10^{-11}J/m$, $K=2\times10^5 J/m^3$, $v_F=5\times10^5m/s$, $\Delta_z=\Delta_{xy}=30meV$. So that: $\delta\approx 7.07nm$, $\lambda=0.672$. We will assume a very small perpendicular anisotropy $K_\perp$, so that it can be neglected compared to the effective anisotropy induced by the TI surface electrons (see next two paragraphs and Appendix \ref{appendix in plane}).

The computation of the effective action for the magnetization is done in Appendixes \ref{appendix in plane},\ref{appendix out of plane}, relaying on the calculations of Sec. \ref{sec effective act electromagnetic}. The total action reads:
\begin{equation}
\Gamma=S_{FM}+\Gamma_{TI}
\label{eq total act}
\end{equation}
with:
\begin{widetext}
\begin{equation}
\Gamma_{TI}=\int dtdxdy\Bigg\{\Delta_{xy}\, \mathbf{s}\cdot\mathbf{m}_{xy}-\frac{d\delta}{2}K^{eff}_\perp\int dx'\, m_x(t,x)\rho_\lambda^2(x)\rho_\lambda^2(x')m_x(t,x')\Bigg\}
\label{eq final eff action}
\end{equation}
and where the spin density $\mathbf{s}$ is:
$$
\mathbf{s}=\frac{\sigma \rho_\lambda^2(x)}{hv_F}\big(\mu-e\Delta V(y)\big)\,\hat{\mathbf{x}}+
$$
\begin{equation}
+\frac{e}{2hv_F}\Bigg(\sigma F_\lambda(x)\,\theta\big(\Delta_z^2\tanh^2(x/\delta)-\mu^2\big)+\frac{\sigma\Delta_z\tanh(x/\delta)}{|\mu|}\,\theta\big(\mu^2-\Delta_z^2\tanh^2(x/\delta)\big)\Bigg)\,\mathbf{E}
\label{eq spin density}
\end{equation}
\end{widetext}
Here $\Delta V(y)=V(y)-V(-L/2)$ is the voltage between $y=-L/2$ and a given point $y$ along the DW, with $V(y)=-E_yy$, and $L$ and $y=\pm L/2$ the length and the end points of the wall respectively. The effective hard axis anisotropy constant $K^{eff}_\perp$ is:
\begin{equation}
K_\perp^{eff}=\frac{\Delta^2_{xy}}{d\delta h v_F}\approx3.49\times10^3J/m^3
\end{equation}
and the functions $\rho_\lambda(x)$ and $F_\lambda(x)$ are given by eqs. (\ref{eq rho}) and (\ref{eq F 1},\ref{eq F 2}) respectively. Remember that the parameter $\lambda$ is given by (see eq. (\ref{eq lambda})):
\begin{equation}
\lambda=\frac{\Delta_z\delta}{\hbar v_F}
\label{eq lambda 2}
\end{equation}
It is important to note also that the spin density is related to the electromagnetic current density as $\mathbf{j}=\pm ev_F\mathbf{s}\times\hat{\mathbf{z}}$.

From eqs. (\ref{eq final eff action},\ref{eq spin density}) we see that there is a spin density that couples to the magnetization, which is related to: 1) (term proportional to $\hat{\mathbf{x}}$ in eq. (\ref{eq spin density})) the edge equilibrium and nonequilibrium currents flowing along the DW, given by eqs. (\ref{eq eq current}) and (\ref{eq noneq current final}) respectively and 2) (term proportional to $\mathbf{E}$ in eq. (\ref{eq spin density})) the topological current due to the anomalous quantum Hall effect in the bulk, coming from the Chern-Simons term (\ref{eq bulk odd}). Besides the spin density there is a non-local contribution induced by the chiral electrons bound to the DW, which acts as a hard axis effective anisotropy energy in the direction perpendicular to the wall.


Although a interfacial DM interaction term of the form $\mathbf{m}\cdot\bm\nabla m_z$ is not induced, as argued in Appendix \ref{appendix out of plane}, the spin density related to the edge current generates a term that resembles it:
\begin{equation}
\Delta_{xy}\, \frac{\sigma \rho_\lambda^2(x)}{hv_F}\big(\mu-e\Delta V(y)\big)\,\hat{\mathbf{x}}\cdot\mathbf{m}_{xy}
\end{equation}
For more clarity, in the special case $\lambda=1$ we can write this as:
\begin{equation}
\Delta_{xy}\, \frac{\mu-e\Delta V(y)}{2hv_F}\,\mathbf{m}_{xy}\cdot\bm\nabla m_z^{(0)}
\end{equation}
while for different values of $\lambda$ small deviations from $\nabla m_z^{(0)}$ take place. The strength of this "pseudo" DM interaction can be tuned by doping with electrons/holes and by the aplication of a voltage between both end points of the DW. Note however that this term is first order in the magnetization, while the DM interaction is second order.

\begin{figure}
\includegraphics[scale=0.8]{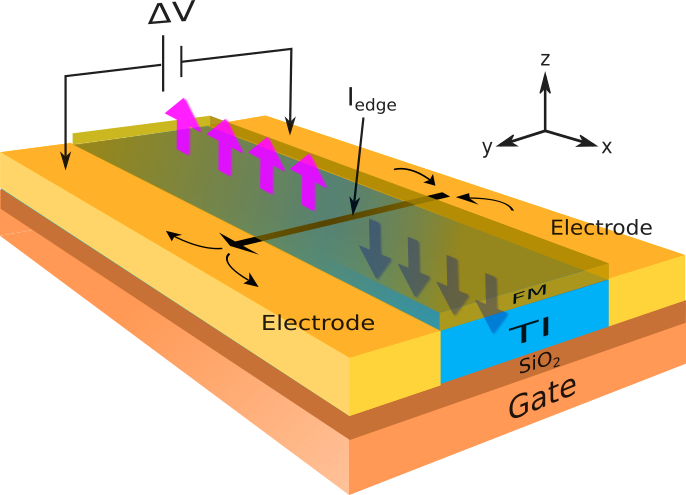}
\caption{\label{fig3} Schematics of a possible experimental setup. The ferromagnetic thin film (green) is deposited on top of the TI (blue) which itself is deposited on top of a gate. Pink and blue thick arrows represent the out-of-plane magnetization pointing up and down respectively. A voltage $\Delta V$ is measured between the electrodes (yellow) at the flanks of the TI. The edge current $I_{edge}$ flowing along the DW is represented by the black arrow connecting both electrodes.}
\end{figure}

In Fig. \ref{fig3} we show the schematics of a possible experimental setup. A thin ferromagnetic film hosting a DW is deposited on top of a TI, which itself is deposited on top of a gate. Two electrodes are attached to the flanks of the TI such that a voltage between both edges is applied, and the gate is used to tune the chemical potential in the TI. A similar setup was realized in\cite{XHW11} where the TI was used as the channel of a field effect transistor.


\section{Phenomenological results}

Now that we have the effective action we can study the phenomenology. We will compute the chirality of the wall in it's equilibrium configuration at finite density and under applied external electric fields. We will also look at the current induced dynamics when a current is applied through the ferromagnet. 

\subsection{Equilibrium configuration of the DW} 

We want to look at minimum energy configurations of the total effective action given by eqs. (\ref{eq total act},\ref{eq final eff action}) of the type: $\mathbf{m}=(sech(x/\delta)\cos\phi,\,sech(x/\delta)\sin\phi,\,\sigma\tanh(x/\delta))$. In principle $\phi$ will be a function of $x$, however to simplify things we will assume $\phi$ to be constant. This will give an average equilibrium value of $\phi(x)$.

From the spin density of eq. (\ref{eq spin density}) we see that an applied electric field in the $x$ direction will not change the average chirality of the wall, since it contributes with terms antisymmetric in $x$ which will vanish when integrated. Contrarily, a voltage in the $y$ direction has a non vanishing effect on the average chirality through the (nonequilibrium) edge current that is generated (term proportional to the voltage in eq. (\ref{eq spin density})). This adds up to the effect of doping the system with electrons/holes, which generates a further (equilibrium) contribution to the edge current (term proportional to the chemical potential in eq. (\ref{eq spin density})).

\begin{figure}
\includegraphics[scale=0.55]{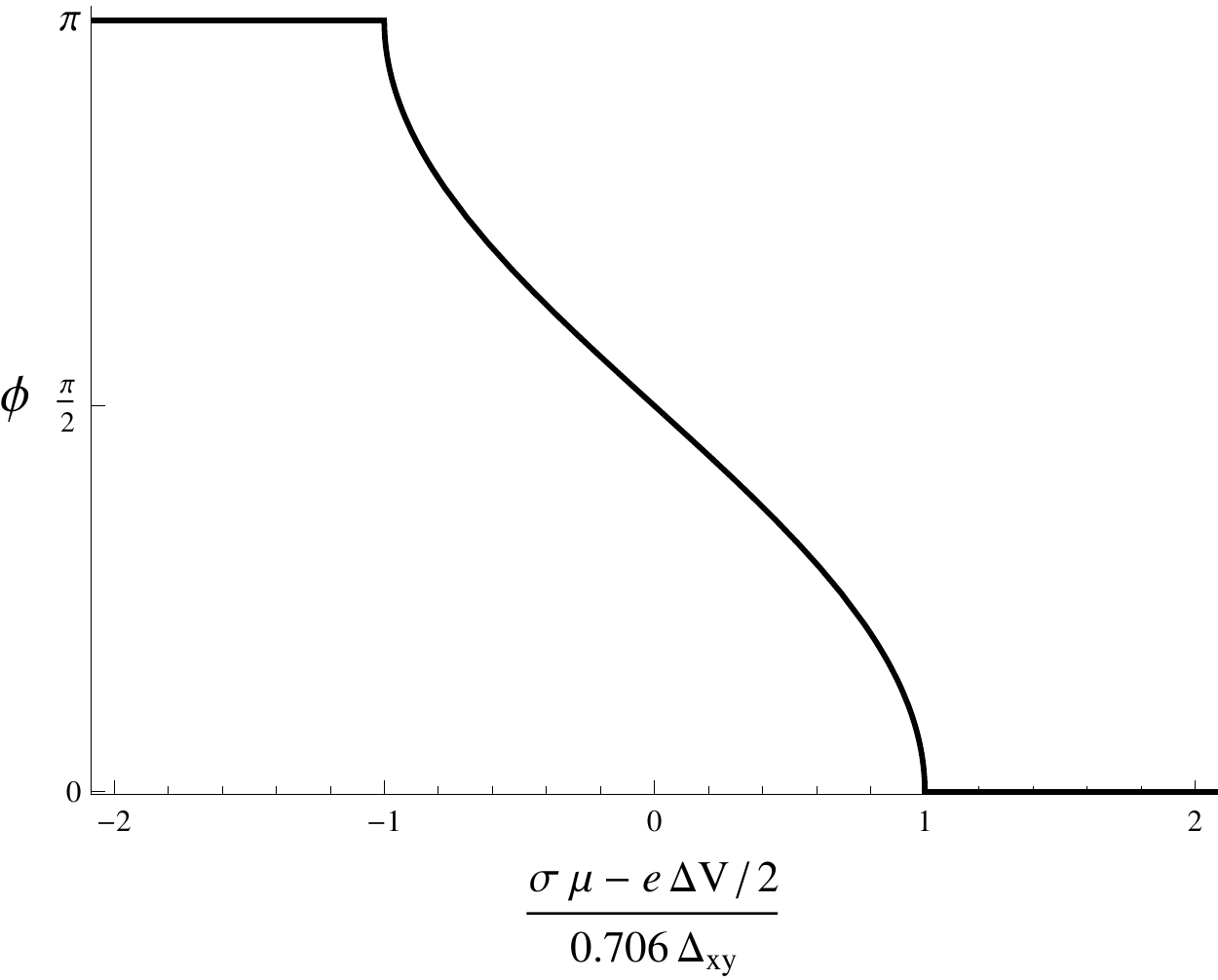}
\caption{\label{fig1} Equilibrium value of the internal angle $\phi$ as a function of the chemical potential and the voltage, for a $\lambda$ parameter value of $\lambda=0.672$.}
\end{figure}

Then at finite density and under an applied voltage in the $y$ direction we obtain the potential energy for the DW (after integration in $x$ and $y$):
\begin{equation}
E(\phi)=\frac{L\Delta_{xy}C_2}{hv_F}\bigg(-C_1\cos\phi+\frac{\Delta_{xy}C_2}{2}\cos^2\phi\bigg)
\end{equation}
with:
\begin{equation}
C_1=\sigma\mu-\frac{e\Delta V}{2}
\end{equation}
\begin{equation}
C_2=-2i\frac{\mathcal{B}_{-1}(1/2+\lambda,-2\lambda)}{\mathcal{B}_{-1}(\lambda,1-2\lambda)}
\end{equation}
where $\Delta V=V(L/2)-V(-L/2)$ is the voltage between both end points of the DW, $\mathcal{B}$ is the incomplete Beta function (see eq. (\ref{eq. beta function})) and $\lambda$ is given in eq. (\ref{eq lambda 2}). Here $C_1$ can take negative values while $C_2$ is real and always positive and fulfills $\lim_{\lambda\rightarrow\infty}C_2=1$. For $|C_1|<\Delta_{xy}C_2$ the energy has a minimum at $\phi=\arccos(C_1/(\Delta_{xy}C_2))$. On the other hand, for $|C_1|\geq \Delta_{xy}C_2$ the minimum energy configuration is $\phi=0$ for $C_1>0$ and $\phi=\pi$ for $C_1<0$. So at $C_1=0$ we have a Bloch DW ($\phi=\pm\pi/2$), and as $|C_1|$ increases $\phi$ is shifted until $|C_1|=\Delta_{xy}C_2$, at which point the DW estabilizes in a \Neel{} configuration, with $\phi=0$ for positive $C_1$ and $\phi=\pi$ for negative $C_1$ (see Fig. \ref{fig1}). This way the chirality of the DW can be tuned by the chemical potential (applying a gate voltage) and the electric field (appliying a voltage between both end points of the DW). Based on this mechanism, there is a way to switch between the two degenerate vacua $\phi=\pm\pi/2$ using an out-of-plane magnetic field to break the degeneracy of the vacuum. It was described in\cite{FC14}, where they made use of an electric field to tune the chirality through the mechanism explained above, while here we have shown that it can be done also by applying a gate voltage.
 
To be specific let us choose the topological charge to be $\sigma=+1$ and switch off the electric field. For the parameters introduced in sec. \ref{section thin film magnet/TI junctions} $\lambda$ takes the value $\lambda=0.672$, so that we have: $C_1=\mu$, $C_2\approx0.706$. For $|\mu|<0.706\Delta_{xy}$ we get $\phi=\arccos(\mu/(0.706\Delta_{xy}))$ while for $|\mu|\geq0.706\Delta_{xy}$ we get $\phi=0$ for positive $\mu$ (TI doped with electrons) and $\phi=\pi$ for negative $\mu$ (TI doped with holes). This behavior qualitatively reproduces the results obtained numerically in\cite{WEN14}. The physical explanation lies in the competition of the torque generated by the chiral edge current on the magnetization and the effective hard axis anisotropy energy, which is also induced by the TI chiral surface electrons.

\subsection{Current induced magnetization dynamics}

Let us apply a current through the ferromagnet in the direction perpendicular to the DW. In terms of the collective coordinates $X(t)$ and $\phi(t)$, where $X$ defines the position of the DW, we can write the total effective action as:
$$
\Gamma=\int dt\bigg(\mathcal{N}\hbar\frac{\dot{X}}{\delta}\phi+\frac{LC_2\Delta_{xy}I_{edge}}{ev_F}\cos\phi-
$$
\begin{equation}
-\frac{LC_2\Delta_{xy}I_{edge}^*}{ev_F}\cos^2\phi-T_{el}\phi-F_{el}X\bigg)
\end{equation}
where $\mathcal{N}=2d\delta L/a^3$ is the number of spins in the DW, the currents $I_{edge}$ and $I_{edge}^*$ are:
\begin{equation}
I_{edge}=\frac{e}{h}\big(\sigma\mu-\frac{e\Delta V}{2}\big)
\end{equation}
\begin{equation}
I_{edge}^*=C_2\frac{e\Delta_{xy}}{2h}\approx4.10\times10^{-7}A
\end{equation}
and $T_{el}$ and $F_{el}$ are the spin-transfer torque and the force generated by the current on the wall, respectively\cite{TKS08,STK11}:
\begin{equation}
T_{el}=\frac{\hbar}{e}I_s\, ;\quad F_{el}=\frac{\hbar}{e\delta}\beta I_s
\end{equation}
Here $I_s=I_{\uparrow}-I_{\downarrow}$ is the applied spin current through the ferromagnet in the positive $x$ direction, and the $\beta$ term is a constant that depends on the microscopic properties of the ferromagnet. It can be quite large for perpendicular anisotropy thin films, where the force from electron reflection can be dominant\cite{STK11}, so following the reference we will fix it to be $\beta=0.3$. 

Now we can write the equations of motion (we include Gilbert damping):
\begin{equation}
e\mathcal{N}\big(\frac{\dot{X}}{\delta}-\alpha\dot{\phi}\big)=\frac{LC_2}{\hbar v_F}(I_{edge}\sin\phi-I_{edge}^*\sin2\phi)-I_s
\end{equation}
\begin{equation}
e\mathcal{N}\big(\dot{\phi}+\alpha\frac{\dot{X}}{\delta}\big)=-\beta\,I_s
\end{equation}
where $\alpha$ is the Gilbert damping parameter, which we will set to be $\alpha=0.01$. From the previous expressions we can obtain a differential equation for $\phi$:
\begin{equation}
\dot{\phi'}=-\frac{j_s}{j_s^*}+\sin2\phi-\frac{I_{edge}}{I_{edge}^*}\sin\phi
\label{eq phi}
\end{equation}
where $I_s=dLj_s$, $I^*_s=dLj^*_s$ and:
\begin{equation}
\phi'=\frac{1+\alpha^2}{\alpha a^3C_2^2}\frac{4\hbar}{K_\perp^{eff}}\,\phi
\end{equation}
\begin{equation}
j_s^*=\frac{e\delta C_2^2}{2\hbar}K_\perp^{eff}\big(\frac{\beta}{\alpha}-1\big)^{-1}\approx3.39\times10^8A/m^2
\end{equation}

For vanishing edge current $I_{edge}$ ($\mu=0$ and $\Delta V=0$) the DW moves with a time-averaged terminal velocity $\langle\dot{X}\rangle=-\beta\delta I_s/(e\alpha \mathcal{N})$ as long as the current flowing through the ferromagnet is smaller than the critical current $I_s^*$. When $I_s$ reaches the value  $I_s^*$ however, Walker breakdown (WB) occurs and $\phi$ starts to change, so that the averaged terminal velocity decreases as $I_s$ increases (see for example\cite{STK11}). 
Thus to obtain high velocities it is important to stay in the non WB regime, which means that one should look for the biggest possible $I_s^*$. We see that it's value is proportional to $\Delta_{xy}^2$, so that the bigger the exchange coupling the higher the velocities that can be achieved. For the actual values of the parameters a velocity $\langle\dot{X}\rangle\approx1.86m/s$ is achieved for an applied current density $j_s=j_s^*$. However if one could achieve an increase of $\Delta_{xy}$ of one order of magnitude ($\Delta_{xy}\sim0.3eV$) the average maximum velocity would increase by two orders of magnitude: $\langle\dot{X}\rangle\approx186m/s$. This way we see that by using a TI, ferromagnets with very week hard axis anisotropy are still suitable for hosting DW motion without suffering from early WB.

\begin{figure}
\includegraphics[scale=0.6]{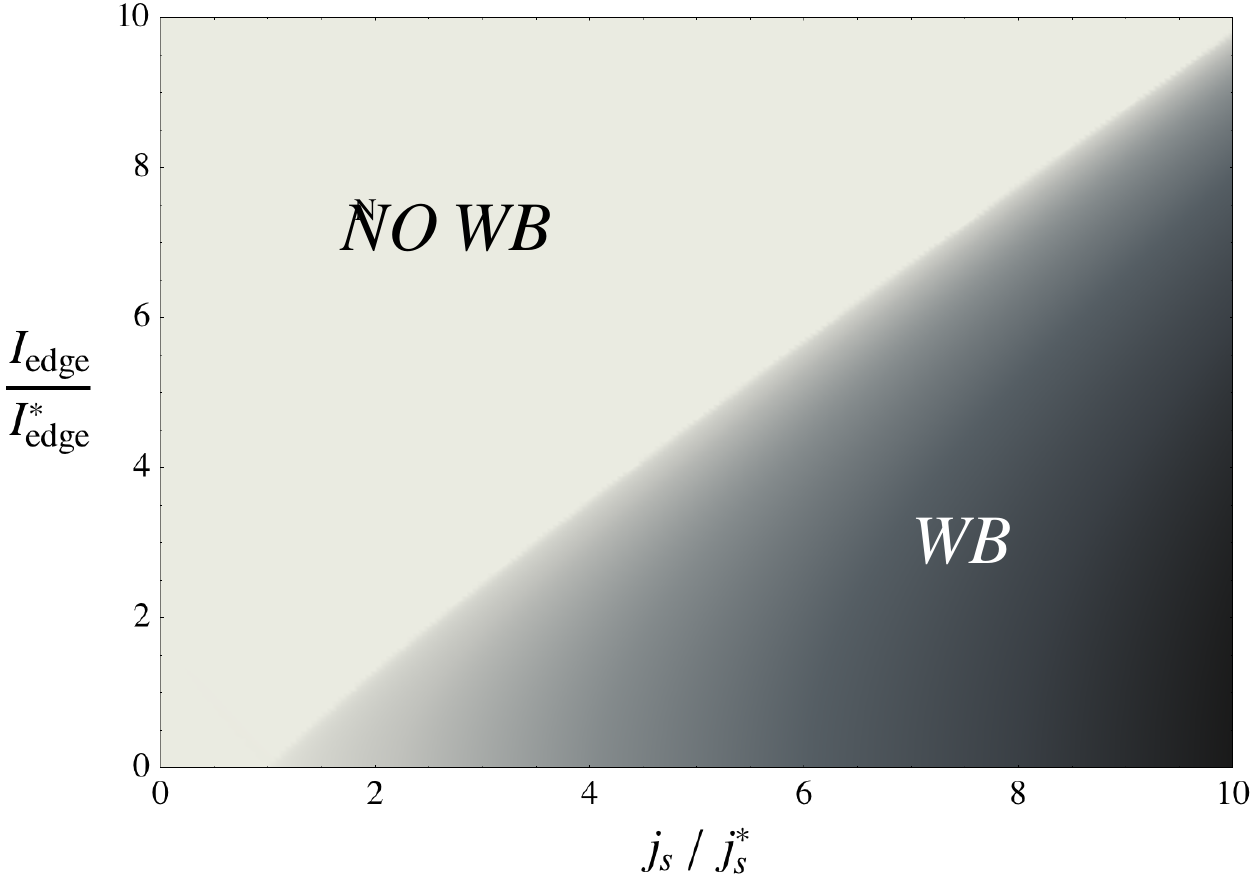}
\caption{\label{fig2} Density plot of $\langle\dot{\phi}\rangle$ in the $I_{edge}/I_{edge}^*-j_s/j_s^*$ plane. Values $\langle\dot{\phi}\rangle<0$ (WB) occur in the dark region, while in the light region $\langle\dot{\phi}\rangle=0$ (no WB). The frontier between the two regions gives the maximum values of $j_s$ before WB occurs.}
\end{figure}

There is a nice way to stabilize the chirality of the DW so that WB is delayed. It was first suggested in\cite{FC14} where an electric field was needed. The chirality stabilization is mediated by the edge current through the third term of the r.h.s. of eq. (\ref{eq phi}). We plot in Fig. \ref{fig2} the values of $j_s$ at which WB occurs for different values of the edge current $I_{edge}$. Notice that DW stabilization can be achieved even in the absence of an electric field by doping the TI with electrons or holes through the application of a gate voltage. Indeed both mechanisms can be added so that higher edge currents can be achieved, and therefore higher terminal velocities can be reached.

\section{Conclusions}

We have analytically computed an approximate effective action up to second order in the electromagnetic field, for a system of Dirac electrons in two spatial dimensions at finite density and with a DW mass term. We have presented a condensed-matter realization of this system, consisting on a ferromagnet hosting a DW (with out-of-plane magnetization domains) coupled via an exchange interaction to the TI surface electrons. There are three relevant contributions to the effective action coming from quantum fluctuations of the fermionic surface states:
\begin{itemize}
\item The first one is linear in the magnetization, which couples to a spin density in the perpendicular direction to the DW. This spin density is related to the edge current of chiral electrons flowing along the DW, which itself can be seen as made of two pieces: an equilibrium current proportional to the chemical potential and a nonequilibrium current proportional to the applied voltage between both end points of the DW.
\item The second contribution is again linear in the magnetization, but now the spin density to which it is coupled is proportional to and has the direction of the applied electric field, and is of opposite signs to either side of the DW. It is related to the topological current generated by the anomalous quantum Hall effect in the bulk.
\item The last contribution is a non-local term quadratic in the magnetization, induced by the chiral electrons, and which acts as an effective hard axis anisotropy energy in the direction perpendicular to the DW.
\end{itemize}


The competition of the torque exerted by the edge current (first contribution) and the effective hard axis anisotropy energy (third contribution) explains the behavior of the chirality of the DW as the chemical potential and/or the voltage between both end points of the wall are modified. The stabilization of the internal angle through the induced effective hard axis anisotropy allows for the motion of the DW with velocity proportional to the applied current through the ferromagnet, especially in the case of ferromagnetic thin films with very weak in-plane anisotropy, which other ways would suffer of a very early WB. The critical current at which WB occurs is proportional to the effective anisotropy energy, which itself is quadratic in the exchange coupling, so that increasing the value of the exchange coupling would result in a significant increase of the maximum possible DW velocity.

Finally, the edge current flowing along the DW has an interesting effect on the wall dynamics. It further stabilizes the internal angle of the DW, which translates in a delay of the appearance of WB. This means the maximum DW velocity can be increased by applying a gate voltage (doping with electrons or holes) and/or applying a voltage between both end points of the DW. The two effects can be combined so that the edge current is further increased and the WB further delayed.

\acknowledgments{Y. Ferreiros gratefully acknowledge A. Cortijo for valuable comments and suggestions. This research is partially supported by the Spanish MECD Grant No. FIS2011-23713 and by Foundation for Fundamental Research on Matter (FOM), which is part of the Netherlands Organisation for Scientific Research (NWO).}

\appendix

\section{Fermionic spectrum}
\label{sec spectrum}

To obtain the fermionic spectrum, we will treat the gauge field as a perturbation and think of the remaining theory (eq. (\ref{eq class act electrodynamics})) as a free theory. For the free theory, the equations of motion can be obtained from eq. (\ref{eq class act electrodynamics}) for vanishing electromagnetic field:
\begin{equation}
\gamma^0\Big(i\,l^\mu_\nu\gamma^\nu\partial_\mu-m\Big)\Psi=0
\label{eq motion}
\end{equation}
Let us take the following representation for the gamma matrices:
\begin{equation}
\gamma^0=\sigma^3,\quad\gamma^{1,2}=-i\,\sigma^{1,2}
\label{eq gamma}
\end{equation}
and write the bispinor $\Psi$ in the basis:
\begin{equation}
\Psi_{R,L}=e^{ip_0x_0}e^{ip_2x_2}\,\Phi_{R,L}(x_1)\mathbf{u}_{R,L}
\label{eq basis 1}
\end{equation}
with:
\begin{equation}
\mathbf{u}_R=\frac{1}{\sqrt{2}}\begin{pmatrix}
1\\
1
\end{pmatrix},\quad
\mathbf{u}_L=\frac{1}{\sqrt{2}}\begin{pmatrix}
1\\
-1
\end{pmatrix}
\end{equation}
so that $\Psi=\Psi_{R}+\Psi_{L}$. With this representation and in this basis, rewriting the first order coupled equations (\ref{eq motion}) as second order decoupled ones we get:
\begin{subequations}
\begin{equation}
\bigg(\partial_1^2+\frac{\lambda(\lambda-\sigma)}{\delta^2}\, sech(x_1/\delta)\bigg)\Phi_R=\Big(\frac{\lambda^2}{\delta^2}+p_2^2-\frac{p_0^2}{v^2}\Big)\Phi_R
\label{eq fermionic states1}
\end{equation}
\begin{equation}
\bigg(\partial_1^2+\frac{\lambda(\lambda+\sigma)}{\delta^2}\, sech(x_1/\delta)\bigg)\Phi_L=\Big(\frac{\lambda^2}{\delta^2}+p_2^2-\frac{p_0^2}{v^2}\Big)\Phi_L
\label{eq fermionic states2}
\end{equation}
\end{subequations}
with $\lambda=m_0\delta/v$. This is nothing but the Schrodinger equation in a modified Poschl-Teller potential (see for example\cite{F98}). Let us write $p_0^2=v^2k^2+v^2p_2^2+m_0^2$, so that the eigenvalues of eqs. (\ref{eq fermionic states1},\ref{eq fermionic states2}) are  $-k^2$. The general solutions of these equations are\cite{F98}:
\begin{widetext}
\begin{equation}
\Phi_R(\sigma,x_1)=\cosh^{\lambda+\frac{1-\sigma}{2}}(x_1/\delta)\bigg\{B(\lambda)\,{_2}F_1\Big[a+\frac{1-\sigma}{4},b+\frac{1-\sigma}{4},\frac{1}{2};-\sinh^2(x_1/\delta)\Big]+
\nonumber
\end{equation}
\begin{equation}
+iC(\lambda)\sinh(x_1/\delta)\,{_2}F_1\Big[a+\frac{1-\sigma}{4}+\frac{1}{2},b+\frac{1-\sigma}{4}+\frac{1}{2},\frac{3}{2};-\sinh^2(x_1/\delta)\Big]\bigg\}
\label{eq general solution}
\end{equation}
\end{widetext}
and $\Phi_L(\sigma,x_1)=\Phi_R(-\sigma,x_1)$. Here $B(\lambda)$ and $C(\lambda)$ are arbitrary constants, ${_2}F_1$ is the hypergeometric function, and the values of $a$ and $b$ are:
\begin{equation}
a=\frac{1}{2}(\lambda+i\delta k),\quad b=\frac{1}{2}(\lambda-i\delta k)
\end{equation}
Note that the first term of the r.h.s. of eq. (\ref{eq general solution}) is even in $x_1$ while the second term is odd. For $k^2>0$ we have the continuum of extended states, which general solution is precisely given by eq. (\ref{eq general solution}). On the other hand, for $k^2<0$ the solutions are bound to the wall, vanishing at $\pm\infty$\cite{F98}, and the energies are quantized.

Let us have a close look at the bound states. If we set $k\rightarrow ik$ we obtain the following conditions for the solutions to be normalizable (for $\sigma=-1$):
\begin{equation}
\begin{array}{l}\Phi_R\longrightarrow \delta k=\lambda-2n\\
\Phi_L\longrightarrow \delta k=\lambda-1-2n\end{array}
\end{equation}
for the even part of eq. (\ref{eq general solution}) and:
\begin{equation}
\begin{array}{l}\Phi_R\longrightarrow \delta k=\lambda-1-2n\\
\Phi_L\longrightarrow \delta k=\lambda-2-2n\end{array}
\end{equation}
for the odd part, where $n=0,1,2,...$ (for $\sigma=1$ one just has to interchange the chiralities: $\Phi_{R,L}\rightarrow\Phi_{L,R}$). From these normalizability conditions and the general solution of eq. (\ref{eq general solution}) we can obtain the bound states. For $\sigma=-1$ we have:
\begin{widetext}
\begin{equation}
n=0\;\left\{\begin{array}{l}\Phi^{(0)}_R=B_0(\lambda)\cosh^{\lambda+1}(x_1/\delta){_2}F_1\Big[\frac{1}{2},\lambda+\frac{1}{2},\frac{1}{2};-\sinh^2(x_1/\delta)\Big]\\
\Phi^{(0)}_L=0\end{array}\right.
\label{eq. chiral mode}
\end{equation}
\begin{equation}
n=1,3,5,...\;\left\{\begin{array}{l}\Phi^{(n)}_R=iC_n(\lambda)\cosh^{\lambda+1}(x_1/\delta)\sinh(x_1/\delta){_2}F_1\Big[a_n+1,b_n+1,\frac{3}{2};-\sinh^2(x_1/\delta)\Big]\\
\Phi^{(n)}_L=B_n(\lambda)\cosh^{\lambda}(x_1/\delta){_2}F_1\Big[a_n+\frac{1}{2},b_n+\frac{1}{2},\frac{1}{2};-\sinh^2(x_1/\delta)\Big]\end{array}\right.
\end{equation}
\begin{equation}
n=2,4,6,...\;\left\{\begin{array}{l}\Phi^{(n)}_R=B_n(\lambda)\cosh^{\lambda+1}(x_1/\delta){_2}F_1\Big[a_n+\frac{1}{2},b_n+\frac{1}{2},\frac{1}{2};-\sinh^2(x_1/\delta)\Big]\\
\Phi^{(n)}_L=iC_n(\lambda)\cosh^{\lambda}(x_1/\delta)\sinh(x_1/\delta){_2}F_1\Big[a_n+1,b_n+1,\frac{3}{2};-\sinh^2(x_1/\delta)\Big]\end{array}\right.
\end{equation}
\end{widetext}
with:
\begin{equation}
a_n=\frac{n}{2},\quad b_n=\lambda-\frac{n}{2}
\end{equation}
and with energies $p_0=-\sigma vp_2$ for $n=0$ and:
\begin{equation}
p_0=\pm\sqrt{\frac{m_0^2}{\lambda^2}n(2\lambda-n)+v^2p_2^2}
\end{equation}
for $n\neq0$ (again, to obtain the solutions for $\sigma=1$ one just has to interchange the chiralities). There is a maximum $n$ before $k^2$ is positive again: $N<\lambda$. Then we see that the number of bound states is given by the largest integer less than $\lambda+1$. Regarding the constants $B_n$ and $C_n$, we impose the solutions to be normalized and obtain for the chiral state $n=0$:
\begin{equation}
B_0(\lambda)=\sqrt{\frac{(-1/4)^\lambda}{\delta \mathcal{B}_{-1}(\lambda,1-2\lambda)}}
\label{eq B0}
\end{equation}
where $\mathcal{B}_{z}(a,b)$ is the incomplete Beta function and can be written as:
\begin{equation}
\mathcal{B}_{z}(a,b)=\int_0^zt^{a-1}(1-t)^{b-1}dt
\label{eq. beta function}
\end{equation}

Finally we can decompose the fermionic field as follows: (for $\sigma=-1$)
\begin{equation}
\Psi=\sum_{n=0}^N\Psi_R^{(n)}(x_0,x_2)\Phi_R^{(n)}(x_1)+\sum_{n=1}^{N}\Psi_L^{(n)}(x_0,x_2)\Phi_L^{(n)}(x_1)+
\nonumber
\end{equation}
\begin{equation}
+\int dk\,\Big(\Psi^{(k)}_R(x_0,x_2)\Phi^{(k)}_R(x_1)+\Psi^{(k)}_L(x_0,x_2)\Phi^{(k)}_L(x_1)\Big)
\label{eq basis}
\end{equation}
with:
\begin{equation}
\Psi_{R,L}^{(n,k)}(x_0,x_2)=\psi^{(n,k)}_{R,L}(x_0,x_2)\mathbf{u}_{R,L}
\end{equation}
Again for $\sigma=1$ one should interchange the chiralities. The summation is over the bound states and the integral is over the continuum of extended states.

\section{Non-adiabatic correction to the CS term}
\label{sec non adiab CS}

Let us write the CS term as:
\begin{equation}
\Pi_{0,o}^{\mu\nu}=-\frac{e^2\sigma F(x_1)}{4\pi v^2}\epsilon^{\mu\rho\nu}\partial_\rho
\end{equation}
To find the non-adiabatic correction $F(x_1)$ we should impose the anomaly cancellation. The gauge variation of the edge theory (eq. (\ref{eq gauge edge theory}))  has to be canceled by the gauge variation of the CS term. With this condition we get the differential eq. for $F$:
\begin{equation}
\partial_1F(x_1)=2\rho_\lambda^2(x_1)
\end{equation}
There is another condition $F$ should fulfil:
\begin{equation}
\lim_{x_1\rightarrow\pm\infty}F(x_1)= \pm1
\end{equation}
so that in the asymptotic limit eq. (\ref{eq CS}) is recovered. Then $F$ can be computed to be:
\begin{equation}
F_\lambda(x_1)=-\delta B_0^2(\lambda)\,sign(x_1)\,\Im\Big[\mathcal{B}_{-y^2}\Big(\frac{1}{2}-\lambda,\frac{1}{2}\Big)\Big]
\label{eq F 1}
\end{equation}
as long as $\lambda\neq1/2,3/2,5/2...$, with $y=\cosh(x_1/\delta)$ and $\mathcal{B}$ the incomplete Beta function defined in eq. (\ref{eq. beta function}). For half integer $\lambda$ we get:
\begin{widetext}
\begin{equation}
\begin{array}{c}F_{1/2}(x_1)=4\delta B_0^2(\frac{1}{2})\arctan\big(\tanh(\frac{x_1}{2\delta})\big)\\
\\
F_{3/2}(x_1)=\delta B_0^2(\frac{3}{2})\left[2\arctan\big(\tanh(\frac{x_1}{2\delta})\big)+sech(\frac{x_1}{\delta})\tanh(\frac{x_1}{\delta})\right]\\
\\
F_{5/2}(x_1)=\frac{\delta}{4}B_0^2(\frac{5}{2})\left[6\arctan\big(\tanh(\frac{x_1}{2\delta})\big)+sech(\frac{x_1}{\delta})\big(3+2sech(\frac{x_1}{\delta})^2\big)\tanh(\frac{x_1}{\delta})\right]\\
\\
\quad\quad\vdots\end{array}
\label{eq F 2}
\end{equation}
\end{widetext}

\section{Even part of the matter contribution in the bulk theory}
\label{sec matter contr bulk}

The matter contribution to the polarization function in the bulk can be written as a sum of an even plus an odd part:
\begin{equation}
\Pi_{matt}^{\mu\nu}=\Pi_{matt,e}^{\mu\nu}+\Pi_{matt,o}^{\mu\nu}
\end{equation}
The even part is a symmetric second rank tensor. Assuming rotational invariance, the most general tensor of this type can be constructed as a linear combination of $\eta_{\mu\nu}$, $p_\mu p_\nu$, $u_\mu u_\nu$ and $p_\mu u_\nu +p_\nu u_\mu$, where $u_\mu =(1,0,0)$ defines the rest frame of the system (see for example\cite{KG06}). Imposing transversality ($p_\mu\Pi_{matt,e}^{\mu\nu}(p)=0$) we obtain the general form for the even contribution:
\begin{equation}
\Pi_{matt,e}^{\mu\nu}(p)=G_1\Big(\eta^{\mu\nu}-\frac{p^\mu p^\nu}{p^2}\Big)+(G_1+G_2)P_\perp^{\mu\nu}
\end{equation}
with $P_\perp^{00}=P_\perp^{0i}=P_\perp^{i0}$ and:
\begin{equation}
P_\perp^{ij}=\delta^{ij}-\frac{p^ip^j}{|\mathbf{p}|^2}
\end{equation}
and where $G_1$ and $G_2$ are scalar functions of $p^0$ and $|\mathbf{p}|$. In the vacuum there is no preferred rest frame, so $u_\mu$ can not appear. In that case, the even part of the polarization function must be proportional to $\eta^{\mu\nu}-p^\mu p^\nu/p^2$ which implies $G_1=-G_2$.

To simplify the computation we will treat the static limit ($p^0=0$) and long wave limit ($\mathbf{p}=0$) separately. In the static limit we have:
$$
\Pi_{matt,e}^{00}=G_{1}(p^0=0)\,;\quad\Pi_{matt,e}^{0i}=\Pi_{matt,e}^{i0}=0
$$
\begin{equation}
\Pi_{matt,e}^{ij}=G_{2}(p^0=0)\Big(\delta^{ij}-\frac{p^ip^j}{|\mathbf{p}|^2}\Big)
\end{equation}
The calculation of $G_1$ and $G_2$ has been done in the context of massive graphene\cite{SS11}. The only difference with the present case is that in graphene there is a multiplicative factor of $4$ that counts the two valleys and the spin degeneracy. We should note that computations for graphene are done with cut-off regularization, which actually breaks gauge invariance. However the matter part is finite and independent of the regularization, so results\cite{SS11} can directly be imported as long as the degeneracy is set to $1$. In the static limit we have:
\begin{widetext}
$$
G_{1}(p^0=0)=\frac{e^2\,\theta(|\mu|-|m|)}{v^2}\Bigg[-\frac{|m|}{4\pi}+\frac{|\mu|}{2\pi}-\frac{|\mu|}{4\pi}\sqrt{1-\frac{4\,p_F^2}{v^2|\mathbf{p}|^2}}\,\theta(|\mathbf{p}|-2p_F)-
$$
\begin{equation}
-\frac{v^2|\mathbf{p}|^2-4m^2}{8\pi v|\mathbf{p}|}\left(\arccos\Big(\frac{2|m|}{\sqrt{v^2|\mathbf{p}|^2+4m^2}}\Big)-\arccos\Big(\frac{2|\mu|}{\sqrt{v^2|\mathbf{p}|^2+4m^2}}\Big)\theta(|\mathbf{p}|-2p_F)\right)\Bigg]
\end{equation}
$$
G_{2}(p^0=0)=\frac{e^2\,\theta(|\mu|-|m|)}{v^2}\Bigg[\frac{|m|}{4\pi}-\frac{|\mu|}{4\pi}\sqrt{1-\frac{4\,p_F^2}{v^2|\mathbf{p}|^2}}\,\theta(|\mathbf{p}|-2p_F)+
$$
\begin{equation}
+\frac{v^2|\mathbf{p}|^2-4m^2}{8\pi v|\mathbf{p}|}\left(\arccos\Big(\frac{2|m|}{\sqrt{v^2|\mathbf{p}|^2+4m^2}}\Big)-\arccos\Big(\frac{2|\mu|}{\sqrt{v^2|\mathbf{p}|^2+4m^2}}\Big)\theta(|\mathbf{p}|-2p_F)\right)\Bigg]
\end{equation}
\end{widetext}
where the Fermi momentum is defined as $vp_F=\sqrt{\mu^2-m^2}$. In configuration space, restoring the $x_1$ dependence of $m$ we have: $vp_F=\sqrt{\mu^2-m_0^2\tanh(x_1/\delta)^2}$. When $|\mu|>m_0$ we are safe to do $\theta(|\mathbf{p}|-2p_F)=0$ for the description of the low energy theory as long as $|\mu|$ doesn't get too close to $m_0$. On the other hand, when $|\mu|<m_0$ we will always have a region in space where $p_F$ is small. However, this region is localized near the DW where the adiabaticity is lost. This means that within the adiabatic approximation we are safe to do $\theta(|\mathbf{p}|-2p_F)=0$ for any value of $\mu$ except when $|\mu|\sim m_0$, in which case corrections proportional to $\theta(|\mathbf{p}|-2p_F)$ would be needed.

Then, supposing $|\mu|$ is below or sufficiently above $m_0$, we can forget about the terms proportional to $\theta(|\mathbf{p}|-2p_F)$ and do a derivative expansion to get:
$$
G_{1}(p^0=0)=\theta(|\mu|-|m|)\,e^2\Bigg(\frac{|\mu|-|m|}{2\pi v^2}+
$$
\begin{equation}
-\frac{|\mathbf{p}|^2}{12\pi|m|}+\mathcal{O}\big(\frac{|\mathbf{p}|^4}{|m|^3}\big)\Bigg)
\end{equation}
\begin{equation}
G_{2}(p^0=0)=\theta(|\mu|-|m|)\left(\frac{e^2|\mathbf{p}|^2}{12\pi|m|}+\mathcal{O}\big(\frac{|\mathbf{p}|^4}{|m|^3}\big)\right)
\end{equation}
This terms are finite in the limit $\mathbf{p}\rightarrow0$.

At the same time, in the long wavelength limit we have:
$$
\Pi_{matt,e}^{00}=0\,;\quad\Pi_{matt,e}^{ij}=G_2(\mathbf{p}=0)\,\delta^{ij}
$$
\begin{equation}
\Pi_{matt,e}^{0i}=\Pi_{matt,e}^{i0}=G_1(\mathbf{p}=0)\frac{vp^i}{p^0}
\end{equation}
Note that we maintain a linear dependence in $p^i$ as we impose spatial homogeneity in the electric and magnetic fields. Now we search for terms linear and quadratic in $p^0$ in $G_{1,2}$ which are finite in the limit $\mathbf{p}\rightarrow0$, so that we go up to second order in a derivative expansion. With this procedure we get rid of non-local terms (negative powers in $\mathbf{p}$). From\cite{SS11} we obtain, in the low energy regime $p_0^2<<m^2$:
\begin{equation}
G_{1}(\mathbf{p}=0)=\theta(|\mu|-|m|)\Bigg(\frac{e^2p_0^2}{12\pi v^2|m|}+\mathcal{O}\big(\frac{p_0^4}{|m|^3}\big)+n.l.t\Bigg)
\label{eq G1 lw}
\end{equation}
\begin{equation}
G_{2}(\mathbf{p}=0)=-\theta(|\mu|-|m|)\Bigg(\frac{e^2p_0^2}{12\pi v^2|m|}+\mathcal{O}\big(\frac{p_0^4}{|m|^3}\big)+n.l.t\Bigg)
\label{eq G2 lw}
\end{equation}
where $n.l.t.$ are non-local terms.

Adding the vacuum and matter contributions (in both static and long wavelength limits) just obtained and neglecting the non-local terms in eqs. (\ref{eq G1 lw},\ref{eq G2 lw}) we arrive to the following expression for the even part of the polarization function (in configuration space):
$$
\Pi_{e}^{00}=-\theta(|m|-|\mu|)\frac{e^2|\bm\partial|^2}{12\pi|m|}+
$$
\begin{equation}
+\theta(|\mu|-|m|)\frac{e^2(|\mu|-|m|)}{2\pi v^2}
\end{equation}
\begin{equation}
\Pi_{e}^{0i}=\Pi_{e}^{i0}=\theta(|m|-|\mu|)\frac{e^2\partial^0\partial^i}{12\pi v|m|}
\end{equation}\begin{equation}
\Pi_{e}^{ij}=-\theta(|m|-|\mu|)\frac{e^2\partial^2}{12\pi v^2|m|}\Big(\delta^{ij}+\frac{v^2\partial^i\partial^j}{\partial^2}\Big)
\end{equation}
Let us note that if we do $p_0\rightarrow0$ in the expressions for the long wavelength limit, the polarization function obtained does not coincide with the one we get if we do $\mathbf{p}\rightarrow0$ in the expressions for the static limit. This is due to the non-locality of the matter part. Withing our approximations, the commutator of the limits $p_0\rightarrow0$ and $\mathbf{p}\rightarrow0$ gives precisely the constant term of $G_{1}(p^0=0)$, which contributes to $\Pi_{e}^{00}$ in the static limit. This means that when adding the static plus the long wavelength contributions, non-local terms which are constant in the limit $p_0\rightarrow0$ and zero in the limit $\mathbf{p}\rightarrow0$ are being approximated by a constant term.

\section{Action for the in-plane magnetization}
\label{appendix in plane}

To obtain the action for the in-plane magnetization we can directly import results from section \ref{sec effective act electromagnetic}. The only place where a bit of care is needed is in the computation of the action for the chiral edge theory, as the sign of the chiral anomaly depends on the sign of the mass. For clarity we will obtain the first order and second order terms (in the magnetization) separately.

\subsection{First order terms}

From eqs. (\ref{action chiral eq},\ref{eq anomaly},\ref{eq bulk even},\ref{eq bulk odd}) we obtain:
\begin{equation}
\Gamma^{(1)}=\int dtdxdy\,\Delta_{xy} \,\mathbf{s}\cdot\mathbf{m}_{xy}
\end{equation}
where the spin density $\mathbf{s}$ is (restoring $\hbar$):
\begin{widetext}
$$
\mathbf{s}=\frac{\rho_\lambda^2(x)}{h}\Bigg(\frac{\sigma\mu}{v_F}\pm \frac{e}{\partial_0\pm\sigma v_F\partial_y}E_y\Bigg)\,\hat{\mathbf{x}}+
$$
\begin{equation}
+\frac{e}{2hv_F}\Bigg(\sigma F_\lambda(x)\,\theta\Big(\Delta_z^2\tanh^2(x/\delta)-\mu^2\Big)+\frac{\sigma\Delta_z\tanh(x/\delta)}{|\mu|}\,\theta\Big(\mu^2-\Delta_z^2\tanh^2(x/\delta)\Big)\Bigg)\,\mathbf{E}
\end{equation}
\end{widetext}
It is related to the electromagnetic current density as: $\mathbf{j}=\pm ev_F\mathbf{s}\times\hat{\mathbf{z}}$. To evaluate the nonequilibrium term, which is non-local (the one with inverse derivatives), we will assume the electromagnetic current to be time independent.  We get\cite{FC14}:
\begin{equation}
\pm \frac{1}{\partial_0\pm\sigma v_F\partial_y}E_y=-\frac{\sigma}{v_F}V(y)
\end{equation}
with $V(y)=-E_y\,y+constant$. We will further set the voltage to zero at $y=-L/2$, where $L$ is the length and $y=\pm L/2$ are the end points of the DW. We finally have:
\begin{equation}
\pm \frac{1}{\partial_0\pm\sigma v_F\partial_y}E_y=-\frac{\sigma}{v_F}\Delta V(y)
\end{equation}
where $\Delta V(y)=V(y)-V(-L/2)$ is the voltage between the end point $y=-L/2$ and a given point $y$ along the wall. The nonequilibrium edge current density along the DW then reads:
\begin{equation}
j^a_{ne}=\pm\frac{e^2\Delta V(y)}{h v_F}\rho^2_\lambda(x)\,(1,\sigma v_F)
\label{eq noneq current final}
\end{equation}
For more clarity let us take the spatial component and compute the average $\langle j^y_{ne}\rangle$:
\begin{equation}
\langle j^y_{ne}\rangle=\frac{1}{L}\int_{-\frac{L}{2}}^{\frac{L}{2}} dy\, j^y_{ne}=\pm\sigma\frac{e^2}{2h}|\rho_\lambda(x)|^2\Delta V
\end{equation}
where here $\Delta V=V(L/2)-V(-L/2)$ is the voltage between both end points of the DW. A similar current configuration has been used recently in the literature\cite{TL12,FC14}.

Finally we can write the spin density as:
\begin{widetext}
$$
\mathbf{s}=\frac{\sigma \rho_\lambda^2(x)}{hv_F}\big(\mu-e\Delta V(y)\big)\,\hat{\mathbf{x}}+
$$
\begin{equation}
+\frac{e}{2hv_F}\Bigg(\sigma F_\lambda(x)\,\theta\big(\Delta_z^2\tanh^2(x/\delta)-\mu^2\big)+\frac{\sigma\Delta_z\tanh(x/\delta)}{|\mu|}\,\theta\big(\mu^2-\Delta_z^2\tanh^2(x/\delta)\big)\Bigg)\,\mathbf{E}
\end{equation}
\end{widetext}

\subsection{Second order terms}
\label{app in plane second order}

The quadratic terms in the magnetization can be obtained from eqs. (\ref{eq anomaly},\ref{eq bulk even},\ref{eq bulk odd}). Expressions (\ref{eq bulk even}) (taken in the static limit) and (\ref{eq bulk odd}) give corrections to the exchange energy and the Berry phase term respectively. Their contribution to the second order terms of the effective action is:
\begin{widetext}
$$
\Gamma^{(2)}_{bulk}=-d\int dtdxdy\,\Bigg\{A^{eff}\theta\big(\mu^2-\Delta_z^2\tanh^2(x/\delta)\big)\,\Big((\partial_xm_x)^2+(\partial_ym_y)^2\Big)\mp
$$
\begin{equation}
\mp\frac{M_s^{eff}}{\gamma}\Big(\sigma F_\lambda(x_1)\,\theta\big(\Delta_z^2\tanh^2(x/\delta)-\mu^2\big)+\frac{\sigma\Delta_z\tanh(x/\delta)}{|\mu|}\theta\big(\mu^2-\Delta_z^2\tanh^2(x/\delta)\big)\Big)\Big(m_y\partial_tm_x-m_x\partial_tm_y\Big)\Bigg\}
\end{equation}
\end{widetext}
with:
\begin{equation}
A^{eff}=\frac{\Delta_{xy}^2}{12\pi\Delta_zd};\quad M_s^{eff}=\frac{\Delta_{xy}^2\gamma}{2hd v_F^2}
\end{equation}
The values of the effective exchange constant and saturation magnetization are $A^{eff}\approx 4.24\times10^{-14}J/m$ and $M_s^{eff}\approx 4.3A/m$, which is much smaller than the values of the ferromagnet $A$ and $M_s$, so that the contribution $\Gamma^{(2)}_{bulk}$ can be neglected. Besides the exchange energy and saturation magnetization renormalization, eq. (\ref{eq bulk even}) gives also an extra dynamical contribution, but is second order in time derivatives, and hence higher order in the derivative expansion than the CS contribution and can be neglected. There would also be further non local dynamical corrections at finite density, which are highly non trivial to compute and which should be again of no importance for the physics compared to the Berry phase term of the ferromagnet.

Regarding the second order contribution coming from the edge theory (eq. (\ref{eq anomaly})) we can write:
$$
\Gamma^{(2)}_{anomaly}=-\frac{\Delta^2_{xy}}{2h v_F}\int dtdt'dxdx'dydy'\rho^2_\lambda(x)\rho^2_\lambda(x')\times
$$
\begin{equation}
\times m_x(t,x,y)(-i\partial_t)G(t-t',y-y')m_x(t',x',y')
\end{equation}
where we defined the green function $G$ as:
$$
G(t-t',y-y')=\int \frac{d^2q}{4\pi^2} \frac{e^{iq_0(t-t')}e^{iq_y(y-y')}}{q_0\mp\sigma v_Fq_y}=
$$
\begin{equation}
=\frac{i}{2}sign(t-t')\delta(\pm\sigma v_F(t-t')+y-y')
\end{equation}
Doing the derivative of the Green function we get:
\begin{widetext}
$$
\Gamma^{(2)}_{anomaly}=-\frac{\Delta^2_{xy}}{2h v_F}\int dtdxdx'dy\,\rho^2_\lambda(x)\rho^2_\lambda(x')\bigg(\,m_x(t,x,y)m_x(t,x',y)+
$$
\begin{equation}
+\frac{1}{2}\int dt'dy'm_x(t,x,y)sign(t-t')\partial_t\delta(\pm\sigma v_F(t-t')+y-y')m_x(t',x',y')\bigg)
\label{Chiral action non local final}
\end{equation}
\end{widetext}

Taking into account that the magnetization does not depend on the $y$ coordinate, the second term of the r.h.s. of eq. (\ref{Chiral action non local final}) can be simplified to:
$$
\pm\sigma\frac{\Delta^2_{xy}}{4h v_F}\int dtdxdx'dy\,\rho^2_\lambda(x)\rho^2_\lambda(x')\, m_x(t,x)\times
$$
\begin{equation}
\times\Big(m_x(t\mp L/v_F,x')+m_x(t\pm L/v_F,x')-2m_x(t,x')\Big)
\end{equation}
where $L$ is the DW length, and $L/v_F$ is the typical time that takes an electron to travel the whole length of the wall. 
Making use of the translation operator we can write:
$$
m_x(t\pm L/v_F,x)=e^{\pm\frac{L}{v_F}\partial_t}m_x(t,x)=
$$
\begin{equation}
=\Big(1\pm\frac{L}{v_F}\partial_t+\frac{L^2}{2v^2_F}\partial^2_t+\mathcal{O}\big(\frac{L^3}{v^3_F}\partial^3_t\big)\Big)m_x(t,x)
\end{equation}
so that we have:
$$
\pm\sigma\frac{\Delta^2_{xy}}{4h v_F}\frac{L^2}{v^2_F}\int dtdxdx'dy\,|\rho_\lambda(x)|^2|\rho_\lambda(x')|^2\times
$$
\begin{equation}
\times\, m_x(t,x)\partial_t^2m_x(t,x')+\mathcal{O}\big(\frac{L^3}{v^3_F}\partial^3_t\big)
\end{equation}
This is second and higher order in derivatives so can be neglected, and we finally get:
$$
\Gamma^{(2)}_{anomaly}=-\frac{d\delta}{2}K_\perp^{eff}\int dtdxdx'dy\,\rho^2_\lambda(x)\rho^2_\lambda(x')\times
$$
\begin{equation}
\times\,m_x(t,x)m_x(t,x')
\end{equation}
where the effective hard axis anisotropy constant is:
\begin{equation}
K_\perp^{eff}=\frac{\Delta^2_{xy}}{d\delta h v_F}\approx3.49\times10^3J/m^3
\label{eq eff haxis anisotropy constant}
\end{equation}
Assuming a ferromagnetic thin film with a weak in-plane anisotropy, the whole contribution to the hard axis anysotropy can be assumed to come from $K_\perp^{eff}$.

\section{Action for the out-of-plane magnetization}
\label{appendix out of plane}

Let us now compute the contributions linear and quadratic in $\tilde{m}_z$. We integrate out the fermions in eq. (\ref{TIact}) with magnetization given by eq. (\ref{eq total mag}), which includes fluctuations around the equilibrium configuration (eq. (\ref{eq equilibrium mag})). We work in the adiabatic approximation taking the mass $m=\pm\sigma\Delta_z\tanh(x/\delta)$ as a constant and at the end restoring the $x$ dependence. Picking only the terms linear and quadratic in $\tilde{m}_z$ we have, in imaginary time:
$$
\Gamma_{z}=-\frac{1}{2v_F^2}\int\frac{d^3p\,d^3q}{(2\pi)^6}\big((q_0-i\mu)^2+\mathcal{E}^2(\mathbf{q})\big)^{-1}\times
$$
\begin{equation}
\times\big((q_0+p_0-i\mu)^2+\mathcal{E}^2(\mathbf{q}+v_F\mathbf{p})\big)^{-1}\big(A+B+C\big)
\label{eq eff act app B 1}
\end{equation}
where $\mathcal{E}(\mathbf{q})=\sqrt{|\mathbf{q}|^2+m^2}$ and:
\begin{equation}
A=-\Delta_z^2\,Tr\Big[\big(i(\slashed{\bar{q}}+\slashed{p})-m\big)\big(i\slashed{\bar{q}}-m\big)\Big]\tilde{m}_z(p)\tilde{m}_z(-p)
\end{equation}
\begin{equation}
B=i\Delta_z\,Tr\Big[\big(i(\slashed{\bar{q}}+\slashed{p})-m\big)\slashed{a}(p)\big(i\slashed{\bar{q}}-m\big)\Big]\tilde{m}_z(-p)
\end{equation}
\begin{equation}
C=i\Delta_z\,Tr\Big[\big(i(\slashed{\bar{q}}+\slashed{p})-m\big)\big(i\slashed{\bar{q}}-m\big)\slashed{a}(-p)\Big]\tilde{m}_z(p)
\end{equation}
We defined $\bar{q}^\mu=(q^0-i\mu,\mathbf{q})$, $p^\mu=(p^0,v_F\mathbf{p})$ and $\slashed{s}=\gamma^\mu s^\nu\delta_{\mu\nu}$ since we are working in euclidean spacetime. It can be seen that at $\mu=0$ the calculation gives no dynamical terms first order in time derivatives, so in this limit dynamical contributions can be neglected. At finite $\mu$ we will take the static limit, assuming any dynamical contribution arising from finite density effects can again be neglected compared to the Berry phase term of the ferromagnet. Hence doing $p_0=0$ and performing the traces we get:
\begin{equation}
A=\Delta_z^2\Big(2\bar{q}_0^2+2|\mathbf{q}|^2-2m^2+\bar{q}_i p_j\delta^{ij}\Big)\tilde{m}_z(p)\tilde{m}_z(-p)
\end{equation}
$$
B=2\Delta_z\Big(\epsilon^{0\alpha\beta}\bar{q}_0p_\alpha a_\beta(p)+\epsilon^{0\alpha\beta}\bar{q}_\alpha p_\beta a_0(p)+
$$
\begin{equation}
+2m\bar{q}_\alpha a_\beta(p)\delta^{\alpha\beta}+m\,\mathbf{p}\cdot\mathbf{a}(p)\Big)\tilde{m}_z(-p)
\end{equation}
$$
C=-2\Delta_z\Big(\epsilon^{0\alpha\beta}\bar{q}_0p_\alpha a_\beta(-p)+\epsilon^{0\alpha\beta}\bar{q}_\alpha p_\beta a_0(-p)-
$$
\begin{equation}
-2m\bar{q}_\alpha a_\beta(-p)\delta^{\alpha\beta}-m\,\mathbf{p}\cdot\mathbf{a}(-p)\Big)\tilde{m}_z(p)
\end{equation}
The first term of $B$ and $C$ has the form of a DM interaction $\mathbf{m}_{xy}\cdot\mathbf{p}\,\tilde{m}_z$, while the second term would give a coupling to the electric field.

We go up only to linear order in the external momentum, so we expand the denominator of eq. (\ref{eq eff act app B 1}) to first order in $\mathbf{p}$ and pick only up to linear terms in the resulting expression. Then performing the integrals in the internal momentum and using dimensional regularization for the divergent integrals present in the vacuum contribution we get (still in imaginary time):
\begin{widetext}
$$
\Gamma_z=\frac{\Delta_z}{16\pi v_F}\int\frac{d^3p}{(2\pi)^3}\frac{m(\mu^2-m^2)}{|\mu|^3}\theta(\mu^2-m^2)\,\mathbf{p}\cdot\Big(\mathbf{a}(p)\tilde{m}_z(-p)+\mathbf{a}(-p)\tilde{m}_z(p)\Big)+
$$
\begin{equation}
+\frac{\Delta_z^2}{8\pi v_F^2}\int\frac{d^3p}{(2\pi)^3}\Bigg(4|m|\theta(m^2-\mu^2)+\Big(|m|+|\mu|+\frac{2m^2}{|\mu|}\Big)\theta(\mu^2-m^2)\Bigg)\tilde{m}_z(p)\tilde{m}_z(-p)
\label{eq eff act app B 2}
\end{equation}
\end{widetext}
The first integral of eq. (\ref{eq eff act app B 2}) above is antisymmetric in $\mathbf{p}$ and vanishes up to a total derivative. The second one gives the final non zero result. We see that the terms mixing $a_\mu$ with $\tilde{m}_z$ vanish, which means that there is no coupling of the out-of-plane magnetization with the electric field to this order and that there is not effective DM interaction. The reason these terms vanish resides in the following vanishing integral:
\begin{equation}
\int_{-\infty}^{\infty}\frac{q_0}{2\pi}\big((q_0-i\mu)^2+\mathcal{E}^2(\mathbf{q})\big)^{-2}(q_0-i\mu)=0
\end{equation}

Going back to real time and configuration space, and doing the substitution $m=\pm\sigma\Delta_z\tanh(x/\delta)$ we get (restoring $\hbar$):
\begin{widetext}
$$
\Gamma_z=-\frac{dK^{eff}}{2}\int dtdxdy\,\Bigg(4\,|\tanh(x/\delta)|\,\theta\big(\Delta_z^2\tanh^2(x/\delta)-\mu^2\big)+
$$
\begin{equation}
+\Big(\frac{|\mu|}{\Delta_z}+|\tanh(x/\delta)|+\frac{2\,\Delta_z\tanh^2(x/\delta)}{|\mu|}\Big)\theta\big(\mu^2-\Delta_z^2\tanh^2(x/\delta)\big)\Bigg)\tilde{m}_z^2
\label{eq out of plane affective action}
\end{equation}
\end{widetext}
which gives just a renormalization of the the easy axis anysotropy energy with:
\begin{equation}
K^{eff}=\frac{\pi\Delta_z^3}{v_F^2dh^2}\approx1.05\times10^3J/m^3
\end{equation}
This is much smaller than the easy axis anysotropy constant of the ferromagnet, so this term can be neglected.

\bibliography{2Delectronsbiblio.bib}

\begin{thebibliography}{75}
\expandafter\ifx\csname natexlab\endcsname\relax\def\natexlab#1{#1}\fi
\expandafter\ifx\csname bibnamefont\endcsname\relax
  \def\bibnamefont#1{#1}\fi
\expandafter\ifx\csname bibfnamefont\endcsname\relax
  \def\bibfnamefont#1{#1}\fi
\expandafter\ifx\csname citenamefont\endcsname\relax
  \def\citenamefont#1{#1}\fi
\expandafter\ifx\csname url\endcsname\relax
  \def\url#1{\texttt{#1}}\fi
\expandafter\ifx\csname urlprefix\endcsname\relax\def\urlprefix{URL }\fi
\providecommand{\bibinfo}[2]{#2}
\providecommand{\eprint}[2][]{\url{#2}}

\bibitem[{\citenamefont{Novoselov et~al.}(2004)\citenamefont{Novoselov, Geim,
  Morozov, Jiang, Zhang, Dubonos, Grigorieva, and Firsov}}]{NGM04}
\bibinfo{author}{\bibfnamefont{K.~S.} \bibnamefont{Novoselov}},
  \bibinfo{author}{\bibfnamefont{A.~K.} \bibnamefont{Geim}},
  \bibinfo{author}{\bibfnamefont{S.~V.} \bibnamefont{Morozov}},
  \bibinfo{author}{\bibfnamefont{D.}~\bibnamefont{Jiang}},
  \bibinfo{author}{\bibfnamefont{Y.}~\bibnamefont{Zhang}},
  \bibinfo{author}{\bibfnamefont{S.~V.} \bibnamefont{Dubonos}},
  \bibinfo{author}{\bibfnamefont{I.~V.} \bibnamefont{Grigorieva}},
  \bibnamefont{and} \bibinfo{author}{\bibfnamefont{A.~A.}
  \bibnamefont{Firsov}}, \bibinfo{journal}{Science}
  \textbf{\bibinfo{volume}{306}}, \bibinfo{pages}{666} (\bibinfo{year}{2004}),
  \eprint{http://www.sciencemag.org/content/306/5696/666.full.pdf},
  \urlprefix\url{http://www.sciencemag.org/content/306/5696/666.abstract}.

\bibitem[{\citenamefont{Castro~Neto et~al.}(2009)\citenamefont{Castro~Neto,
  Guinea, Peres, Novoselov, and Geim}}]{NGP09}
\bibinfo{author}{\bibfnamefont{A.~H.} \bibnamefont{Castro~Neto}},
  \bibinfo{author}{\bibfnamefont{F.}~\bibnamefont{Guinea}},
  \bibinfo{author}{\bibfnamefont{N.~M.~R.} \bibnamefont{Peres}},
  \bibinfo{author}{\bibfnamefont{K.~S.} \bibnamefont{Novoselov}},
  \bibnamefont{and} \bibinfo{author}{\bibfnamefont{A.~K.} \bibnamefont{Geim}},
  \bibinfo{journal}{Rev. Mod. Phys.} \textbf{\bibinfo{volume}{81}},
  \bibinfo{pages}{109} (\bibinfo{year}{2009}),
  \urlprefix\url{http://link.aps.org/doi/10.1103/RevModPhys.81.109}.

\bibitem[{\citenamefont{Katsnelson}(2012)}]{K12}
\bibinfo{author}{\bibfnamefont{M.~I.} \bibnamefont{Katsnelson}},
  \bibinfo{journal}{Cambridge University Press}  (\bibinfo{year}{2012}).

\bibitem[{\citenamefont{Fu et~al.}(2007)\citenamefont{Fu, Kane, and
  Mele}}]{FKM07}
\bibinfo{author}{\bibfnamefont{L.}~\bibnamefont{Fu}},
  \bibinfo{author}{\bibfnamefont{C.~L.} \bibnamefont{Kane}}, \bibnamefont{and}
  \bibinfo{author}{\bibfnamefont{E.~J.} \bibnamefont{Mele}},
  \bibinfo{journal}{Phys. Rev. Lett.} \textbf{\bibinfo{volume}{98}},
  \bibinfo{pages}{106803} (\bibinfo{year}{2007}),
  \urlprefix\url{http://link.aps.org/doi/10.1103/PhysRevLett.98.106803}.

\bibitem[{\citenamefont{Hasan and Kane}(2010)}]{HK10}
\bibinfo{author}{\bibfnamefont{M.~Z.} \bibnamefont{Hasan}} \bibnamefont{and}
  \bibinfo{author}{\bibfnamefont{C.~L.} \bibnamefont{Kane}},
  \bibinfo{journal}{Rev. Mod. Phys.} \textbf{\bibinfo{volume}{82}},
  \bibinfo{pages}{3045} (\bibinfo{year}{2010}),
  \urlprefix\url{http://link.aps.org/doi/10.1103/RevModPhys.82.3045}.

\bibitem[{\citenamefont{Qi and Zhang}(2011)}]{XZ11}
\bibinfo{author}{\bibfnamefont{X.-L.} \bibnamefont{Qi}} \bibnamefont{and}
  \bibinfo{author}{\bibfnamefont{S.-C.} \bibnamefont{Zhang}},
  \bibinfo{journal}{Rev. Mod. Phys.} \textbf{\bibinfo{volume}{83}},
  \bibinfo{pages}{1057} (\bibinfo{year}{2011}),
  \urlprefix\url{http://link.aps.org/doi/10.1103/RevModPhys.83.1057}.

\bibitem[{\citenamefont{Parkin et~al.}(2008)\citenamefont{Parkin, Hayashi, and
  Thomas}}]{PHM08}
\bibinfo{author}{\bibfnamefont{S.~S.~P.} \bibnamefont{Parkin}},
  \bibinfo{author}{\bibfnamefont{M.}~\bibnamefont{Hayashi}}, \bibnamefont{and}
  \bibinfo{author}{\bibfnamefont{L.}~\bibnamefont{Thomas}},
  \bibinfo{journal}{Science} \textbf{\bibinfo{volume}{320}},
  \bibinfo{pages}{190} (\bibinfo{year}{2008}).

\bibitem[{\citenamefont{Thomas et~al.}(2010)\citenamefont{Thomas, Moriya,
  Rettner, and Parkin}}]{TMR10}
\bibinfo{author}{\bibfnamefont{L.}~\bibnamefont{Thomas}},
  \bibinfo{author}{\bibfnamefont{R.}~\bibnamefont{Moriya}},
  \bibinfo{author}{\bibfnamefont{C.}~\bibnamefont{Rettner}}, \bibnamefont{and}
  \bibinfo{author}{\bibfnamefont{S.~S.} \bibnamefont{Parkin}},
  \bibinfo{journal}{Science} \textbf{\bibinfo{volume}{330}},
  \bibinfo{pages}{1810} (\bibinfo{year}{2010}).

\bibitem[{\citenamefont{Schryer and Walker}(1974)}]{SW74}
\bibinfo{author}{\bibfnamefont{N.~L.} \bibnamefont{Schryer}} \bibnamefont{and}
  \bibinfo{author}{\bibfnamefont{L.~R.} \bibnamefont{Walker}},
  \bibinfo{journal}{Journal of Applied Physics} \textbf{\bibinfo{volume}{45}}
  (\bibinfo{year}{1974}).

\bibitem[{\citenamefont{Ono et~al.}(1999)\citenamefont{Ono, Miyajima, Shigeto,
  Mibu, Hosoito, and Shinjo}}]{OMS99}
\bibinfo{author}{\bibfnamefont{T.}~\bibnamefont{Ono}},
  \bibinfo{author}{\bibfnamefont{H.}~\bibnamefont{Miyajima}},
  \bibinfo{author}{\bibfnamefont{K.}~\bibnamefont{Shigeto}},
  \bibinfo{author}{\bibfnamefont{K.}~\bibnamefont{Mibu}},
  \bibinfo{author}{\bibfnamefont{N.}~\bibnamefont{Hosoito}}, \bibnamefont{and}
  \bibinfo{author}{\bibfnamefont{T.}~\bibnamefont{Shinjo}},
  \bibinfo{journal}{Science} \textbf{\bibinfo{volume}{284}},
  \bibinfo{pages}{468} (\bibinfo{year}{1999}),
  \eprint{http://www.sciencemag.org/content/284/5413/468.full.pdf},
  \urlprefix\url{http://www.sciencemag.org/content/284/5413/468.abstract}.

\bibitem[{\citenamefont{Atkinson et~al.}(2003)\citenamefont{Atkinson, Allwood,
  Xiong, Cooke, Faulkner, and Cowburn}}]{AAX03}
\bibinfo{author}{\bibfnamefont{D.}~\bibnamefont{Atkinson}},
  \bibinfo{author}{\bibfnamefont{D.~A.} \bibnamefont{Allwood}},
  \bibinfo{author}{\bibfnamefont{G.}~\bibnamefont{Xiong}},
  \bibinfo{author}{\bibfnamefont{M.~D.} \bibnamefont{Cooke}},
  \bibinfo{author}{\bibfnamefont{C.~C.} \bibnamefont{Faulkner}},
  \bibnamefont{and} \bibinfo{author}{\bibfnamefont{R.~P.}
  \bibnamefont{Cowburn}}, \bibinfo{journal}{Nat Mater}
  \textbf{\bibinfo{volume}{2}}, \bibinfo{pages}{85} (\bibinfo{year}{2003}),
  \urlprefix\url{http://dx.doi.org/10.1038/nmat803}.

\bibitem[{\citenamefont{Nakatani et~al.}(2001)\citenamefont{Nakatani, Hayashi,
  Ono, and Miyajima}}]{NHO01}
\bibinfo{author}{\bibfnamefont{Y.}~\bibnamefont{Nakatani}},
  \bibinfo{author}{\bibfnamefont{N.}~\bibnamefont{Hayashi}},
  \bibinfo{author}{\bibfnamefont{T.}~\bibnamefont{Ono}}, \bibnamefont{and}
  \bibinfo{author}{\bibfnamefont{H.}~\bibnamefont{Miyajima}},
  \bibinfo{journal}{Magnetics, IEEE Transactions on}
  \textbf{\bibinfo{volume}{37}}, \bibinfo{pages}{2129} (\bibinfo{year}{2001}),
  ISSN \bibinfo{issn}{0018-9464}.

\bibitem[{\citenamefont{Thiaville et~al.}(2002)\citenamefont{Thiaville,
  Garcı́a, and Miltat}}]{TGM02}
\bibinfo{author}{\bibfnamefont{A.}~\bibnamefont{Thiaville}},
  \bibinfo{author}{\bibfnamefont{J.}~\bibnamefont{Garcı́a}},
  \bibnamefont{and} \bibinfo{author}{\bibfnamefont{J.}~\bibnamefont{Miltat}},
  \bibinfo{journal}{Journal of Magnetism and Magnetic Materials}
  \textbf{\bibinfo{volume}{242–245, Part 2}}, \bibinfo{pages}{1061 }
  (\bibinfo{year}{2002}), ISSN \bibinfo{issn}{0304-8853},
  \bibinfo{note}{proceedings of the Joint European Magnetic Symposia
  (JEMS'01)},
  \urlprefix\url{http://www.sciencedirect.com/science/article/pii/S0304885301013531}.

\bibitem[{\citenamefont{Nakatani et~al.}(2003)\citenamefont{Nakatani,
  Thiaville, and Miltat}}]{NTM03}
\bibinfo{author}{\bibfnamefont{Y.}~\bibnamefont{Nakatani}},
  \bibinfo{author}{\bibfnamefont{A.}~\bibnamefont{Thiaville}},
  \bibnamefont{and} \bibinfo{author}{\bibfnamefont{J.}~\bibnamefont{Miltat}},
  \bibinfo{journal}{Nat Mater} \textbf{\bibinfo{volume}{2}},
  \bibinfo{pages}{521} (\bibinfo{year}{2003}),
  \urlprefix\url{http://dx.doi.org/10.1038/nmat931}.

\bibitem[{\citenamefont{Beach et~al.}(2005{\natexlab{a}})\citenamefont{Beach,
  Nistor, Knutson, Tsoi, and Erskine}}]{BNK05}
\bibinfo{author}{\bibfnamefont{G.~S.~D.} \bibnamefont{Beach}},
  \bibinfo{author}{\bibfnamefont{C.}~\bibnamefont{Nistor}},
  \bibinfo{author}{\bibfnamefont{C.}~\bibnamefont{Knutson}},
  \bibinfo{author}{\bibfnamefont{M.}~\bibnamefont{Tsoi}}, \bibnamefont{and}
  \bibinfo{author}{\bibfnamefont{J.~L.} \bibnamefont{Erskine}},
  \bibinfo{journal}{Nat Mater} \textbf{\bibinfo{volume}{4}},
  \bibinfo{pages}{741} (\bibinfo{year}{2005}{\natexlab{a}}),
  \urlprefix\url{http://dx.doi.org/10.1038/nmat1477}.

\bibitem[{\citenamefont{Berger}(1984)}]{B84}
\bibinfo{author}{\bibfnamefont{L.}~\bibnamefont{Berger}},
  \bibinfo{journal}{Journal of Applied Physics} \textbf{\bibinfo{volume}{55}}
  (\bibinfo{year}{1984}).

\bibitem[{\citenamefont{Freitas and Berger}(1985)}]{FB85}
\bibinfo{author}{\bibfnamefont{P.~P.} \bibnamefont{Freitas}} \bibnamefont{and}
  \bibinfo{author}{\bibfnamefont{L.}~\bibnamefont{Berger}},
  \bibinfo{journal}{Journal of Applied Physics} \textbf{\bibinfo{volume}{57}}
  (\bibinfo{year}{1985}).

\bibitem[{\citenamefont{Hung and Berger}(1988)}]{HB88}
\bibinfo{author}{\bibfnamefont{C.}~\bibnamefont{Hung}} \bibnamefont{and}
  \bibinfo{author}{\bibfnamefont{L.}~\bibnamefont{Berger}},
  \bibinfo{journal}{Journal of Applied Physics} \textbf{\bibinfo{volume}{63}}
  (\bibinfo{year}{1988}).

\bibitem[{\citenamefont{Gan et~al.}(2000)\citenamefont{Gan, Chung, Aschenbach,
  Dreyer, and Gomez}}]{GCA00}
\bibinfo{author}{\bibfnamefont{L.}~\bibnamefont{Gan}},
  \bibinfo{author}{\bibfnamefont{S.}~\bibnamefont{Chung}},
  \bibinfo{author}{\bibfnamefont{K.}~\bibnamefont{Aschenbach}},
  \bibinfo{author}{\bibfnamefont{M.}~\bibnamefont{Dreyer}}, \bibnamefont{and}
  \bibinfo{author}{\bibfnamefont{R.}~\bibnamefont{Gomez}},
  \bibinfo{journal}{Magnetics, IEEE Transactions on}
  \textbf{\bibinfo{volume}{36}}, \bibinfo{pages}{3047} (\bibinfo{year}{2000}),
  ISSN \bibinfo{issn}{0018-9464}.

\bibitem[{\citenamefont{Koo et~al.}(2002)\citenamefont{Koo, Krafft, and
  Gomez}}]{KKG02}
\bibinfo{author}{\bibfnamefont{H.}~\bibnamefont{Koo}},
  \bibinfo{author}{\bibfnamefont{C.}~\bibnamefont{Krafft}}, \bibnamefont{and}
  \bibinfo{author}{\bibfnamefont{R.~D.} \bibnamefont{Gomez}},
  \bibinfo{journal}{Applied Physics Letters} \textbf{\bibinfo{volume}{81}}
  (\bibinfo{year}{2002}).

\bibitem[{\citenamefont{Yamaguchi et~al.}(2004)\citenamefont{Yamaguchi, Ono,
  Nasu, Miyake, Mibu, and Shinjo}}]{YON04}
\bibinfo{author}{\bibfnamefont{A.}~\bibnamefont{Yamaguchi}},
  \bibinfo{author}{\bibfnamefont{T.}~\bibnamefont{Ono}},
  \bibinfo{author}{\bibfnamefont{S.}~\bibnamefont{Nasu}},
  \bibinfo{author}{\bibfnamefont{K.}~\bibnamefont{Miyake}},
  \bibinfo{author}{\bibfnamefont{K.}~\bibnamefont{Mibu}}, \bibnamefont{and}
  \bibinfo{author}{\bibfnamefont{T.}~\bibnamefont{Shinjo}},
  \bibinfo{journal}{Phys. Rev. Lett.} \textbf{\bibinfo{volume}{92}},
  \bibinfo{pages}{077205} (\bibinfo{year}{2004}),
  \urlprefix\url{http://link.aps.org/doi/10.1103/PhysRevLett.92.077205}.

\bibitem[{\citenamefont{Grollier et~al.}(2003)\citenamefont{Grollier, Boulenc,
  Cros, Hamzić, Vaurès, Fert, and Faini}}]{GBC03}
\bibinfo{author}{\bibfnamefont{J.}~\bibnamefont{Grollier}},
  \bibinfo{author}{\bibfnamefont{P.}~\bibnamefont{Boulenc}},
  \bibinfo{author}{\bibfnamefont{V.}~\bibnamefont{Cros}},
  \bibinfo{author}{\bibfnamefont{A.}~\bibnamefont{Hamzić}},
  \bibinfo{author}{\bibfnamefont{A.}~\bibnamefont{Vaurès}},
  \bibinfo{author}{\bibfnamefont{A.}~\bibnamefont{Fert}}, \bibnamefont{and}
  \bibinfo{author}{\bibfnamefont{G.}~\bibnamefont{Faini}},
  \bibinfo{journal}{Applied Physics Letters} \textbf{\bibinfo{volume}{83}}
  (\bibinfo{year}{2003}).

\bibitem[{\citenamefont{Yan et~al.}(2011)\citenamefont{Yan, Wang, and
  Wang}}]{YWW11}
\bibinfo{author}{\bibfnamefont{P.}~\bibnamefont{Yan}},
  \bibinfo{author}{\bibfnamefont{X.~S.} \bibnamefont{Wang}}, \bibnamefont{and}
  \bibinfo{author}{\bibfnamefont{X.~R.} \bibnamefont{Wang}},
  \bibinfo{journal}{Phys. Rev. Lett.} \textbf{\bibinfo{volume}{107}},
  \bibinfo{pages}{177207} (\bibinfo{year}{2011}),
  \urlprefix\url{http://link.aps.org/doi/10.1103/PhysRevLett.107.177207}.

\bibitem[{\citenamefont{Kim et~al.}(2012)\citenamefont{Kim, St\"ark, Kl\"aui,
  Yoon, You, Lopez-Diaz, and Martinez}}]{KSK12}
\bibinfo{author}{\bibfnamefont{J.-S.} \bibnamefont{Kim}},
  \bibinfo{author}{\bibfnamefont{M.}~\bibnamefont{St\"ark}},
  \bibinfo{author}{\bibfnamefont{M.}~\bibnamefont{Kl\"aui}},
  \bibinfo{author}{\bibfnamefont{J.}~\bibnamefont{Yoon}},
  \bibinfo{author}{\bibfnamefont{C.-Y.} \bibnamefont{You}},
  \bibinfo{author}{\bibfnamefont{L.}~\bibnamefont{Lopez-Diaz}},
  \bibnamefont{and} \bibinfo{author}{\bibfnamefont{E.}~\bibnamefont{Martinez}},
  \bibinfo{journal}{Phys. Rev. B} \textbf{\bibinfo{volume}{85}},
  \bibinfo{pages}{174428} (\bibinfo{year}{2012}),
  \urlprefix\url{http://link.aps.org/doi/10.1103/PhysRevB.85.174428}.

\bibitem[{\citenamefont{Wang et~al.}(2012)\citenamefont{Wang, Guo, Nie, Zhang,
  and Li}}]{WGN12}
\bibinfo{author}{\bibfnamefont{X.-g.} \bibnamefont{Wang}},
  \bibinfo{author}{\bibfnamefont{G.-h.} \bibnamefont{Guo}},
  \bibinfo{author}{\bibfnamefont{Y.-z.} \bibnamefont{Nie}},
  \bibinfo{author}{\bibfnamefont{G.-f.} \bibnamefont{Zhang}}, \bibnamefont{and}
  \bibinfo{author}{\bibfnamefont{Z.-x.} \bibnamefont{Li}},
  \bibinfo{journal}{Phys. Rev. B} \textbf{\bibinfo{volume}{86}},
  \bibinfo{pages}{054445} (\bibinfo{year}{2012}),
  \urlprefix\url{http://link.aps.org/doi/10.1103/PhysRevB.86.054445}.

\bibitem[{\citenamefont{Wang et~al.}(2013)\citenamefont{Wang, Guo, Zhang, Nie,
  and Xia}}]{WGZ13}
\bibinfo{author}{\bibfnamefont{X.-g.} \bibnamefont{Wang}},
  \bibinfo{author}{\bibfnamefont{G.-h.} \bibnamefont{Guo}},
  \bibinfo{author}{\bibfnamefont{G.-f.} \bibnamefont{Zhang}},
  \bibinfo{author}{\bibfnamefont{Y.-z.} \bibnamefont{Nie}}, \bibnamefont{and}
  \bibinfo{author}{\bibfnamefont{Q.-l.} \bibnamefont{Xia}},
  \bibinfo{journal}{Applied Physics Letters} \textbf{\bibinfo{volume}{102}},
  \bibinfo{eid}{132401} (\bibinfo{year}{2013}),
  \urlprefix\url{http://scitation.aip.org/content/aip/journal/apl/102/13/10.1063/1.4799285}.

\bibitem[{\citenamefont{Hata et~al.}(2014)\citenamefont{Hata, Taniguchi, Lee,
  Moriyama, and Ono}}]{HTH14}
\bibinfo{author}{\bibfnamefont{H.}~\bibnamefont{Hata}},
  \bibinfo{author}{\bibfnamefont{T.}~\bibnamefont{Taniguchi}},
  \bibinfo{author}{\bibfnamefont{H.-W.} \bibnamefont{Lee}},
  \bibinfo{author}{\bibfnamefont{T.}~\bibnamefont{Moriyama}}, \bibnamefont{and}
  \bibinfo{author}{\bibfnamefont{T.}~\bibnamefont{Ono}},
  \bibinfo{journal}{Applied Physics Express} \textbf{\bibinfo{volume}{7}},
  \bibinfo{pages}{033001} (\bibinfo{year}{2014}),
  \urlprefix\url{http://stacks.iop.org/1882-0786/7/i=3/a=033001}.

\bibitem[{\citenamefont{Manchon et~al.}(2014)\citenamefont{Manchon, Ndiaye,
  Moon, Lee, and Lee}}]{MNM14}
\bibinfo{author}{\bibfnamefont{A.}~\bibnamefont{Manchon}},
  \bibinfo{author}{\bibfnamefont{P.~B.} \bibnamefont{Ndiaye}},
  \bibinfo{author}{\bibfnamefont{J.-H.} \bibnamefont{Moon}},
  \bibinfo{author}{\bibfnamefont{H.-W.} \bibnamefont{Lee}}, \bibnamefont{and}
  \bibinfo{author}{\bibfnamefont{K.-J.} \bibnamefont{Lee}},
  \bibinfo{journal}{Phys. Rev. B} \textbf{\bibinfo{volume}{90}},
  \bibinfo{pages}{224403} (\bibinfo{year}{2014}),
  \urlprefix\url{http://link.aps.org/doi/10.1103/PhysRevB.90.224403}.

\bibitem[{\citenamefont{Wang et~al.}(2015)\citenamefont{Wang, Albert, Beg,
  Bisotti, Chernyshenko, Cort\'es-Ortu\~no, Hawke, and Fangohr}}]{WAB15}
\bibinfo{author}{\bibfnamefont{W.}~\bibnamefont{Wang}},
  \bibinfo{author}{\bibfnamefont{M.}~\bibnamefont{Albert}},
  \bibinfo{author}{\bibfnamefont{M.}~\bibnamefont{Beg}},
  \bibinfo{author}{\bibfnamefont{M.-A.} \bibnamefont{Bisotti}},
  \bibinfo{author}{\bibfnamefont{D.}~\bibnamefont{Chernyshenko}},
  \bibinfo{author}{\bibfnamefont{D.}~\bibnamefont{Cort\'es-Ortu\~no}},
  \bibinfo{author}{\bibfnamefont{I.}~\bibnamefont{Hawke}}, \bibnamefont{and}
  \bibinfo{author}{\bibfnamefont{H.}~\bibnamefont{Fangohr}},
  \bibinfo{journal}{Phys. Rev. Lett.} \textbf{\bibinfo{volume}{114}},
  \bibinfo{pages}{087203} (\bibinfo{year}{2015}),
  \urlprefix\url{http://link.aps.org/doi/10.1103/PhysRevLett.114.087203}.

\bibitem[{\citenamefont{Onho et~al.}(2000)\citenamefont{Onho, Chiba, Matsukura,
  Oyima, Abe, Dietl, Ohno, and Ohtani}}]{OCM00}
\bibinfo{author}{\bibfnamefont{H.}~\bibnamefont{Onho}},
  \bibinfo{author}{\bibfnamefont{D.}~\bibnamefont{Chiba}},
  \bibinfo{author}{\bibfnamefont{F.}~\bibnamefont{Matsukura}},
  \bibinfo{author}{\bibfnamefont{T.}~\bibnamefont{Oyima}},
  \bibinfo{author}{\bibfnamefont{E.}~\bibnamefont{Abe}},
  \bibinfo{author}{\bibfnamefont{T.}~\bibnamefont{Dietl}},
  \bibinfo{author}{\bibfnamefont{Y.}~\bibnamefont{Ohno}}, \bibnamefont{and}
  \bibinfo{author}{\bibfnamefont{K.}~\bibnamefont{Ohtani}},
  \bibinfo{journal}{Nature} \textbf{\bibinfo{volume}{408}},
  \bibinfo{pages}{944} (\bibinfo{year}{2000}).

\bibitem[{\citenamefont{Schellekens et~al.}(2011)\citenamefont{Schellekens,
  van~den Brink, Franken, Swagten, and Koopmans}}]{SBF11}
\bibinfo{author}{\bibfnamefont{A.}~\bibnamefont{Schellekens}},
  \bibinfo{author}{\bibfnamefont{A.}~\bibnamefont{van~den Brink}},
  \bibinfo{author}{\bibfnamefont{J.}~\bibnamefont{Franken}},
  \bibinfo{author}{\bibfnamefont{H.}~\bibnamefont{Swagten}}, \bibnamefont{and}
  \bibinfo{author}{\bibfnamefont{B.}~\bibnamefont{Koopmans}},
  \bibinfo{journal}{Nat. Commun.} \textbf{\bibinfo{volume}{3}},
  \bibinfo{pages}{847} (\bibinfo{year}{2011}).

\bibitem[{\citenamefont{Beach et~al.}(2005{\natexlab{b}})\citenamefont{Beach,
  Nistor, Knutson, Tsoi, and Erskine}}]{GCC05}
\bibinfo{author}{\bibfnamefont{G.~S.~D.} \bibnamefont{Beach}},
  \bibinfo{author}{\bibfnamefont{C.}~\bibnamefont{Nistor}},
  \bibinfo{author}{\bibfnamefont{C.}~\bibnamefont{Knutson}},
  \bibinfo{author}{\bibfnamefont{M.}~\bibnamefont{Tsoi}}, \bibnamefont{and}
  \bibinfo{author}{\bibfnamefont{J.~L.} \bibnamefont{Erskine}},
  \bibinfo{journal}{Nat Mater} \textbf{\bibinfo{volume}{4}},
  \bibinfo{pages}{741} (\bibinfo{year}{2005}{\natexlab{b}}).

\bibitem[{\citenamefont{Brataas}(2013)}]{B13}
\bibinfo{author}{\bibfnamefont{A.}~\bibnamefont{Brataas}},
  \bibinfo{journal}{Nat Nano} \textbf{\bibinfo{volume}{8}},
  \bibinfo{pages}{485} (\bibinfo{year}{2013}),
  \urlprefix\url{http://dx.doi.org/10.1038/nnano.2013.126}.

\bibitem[{\citenamefont{Thiaville et~al.}(2012)\citenamefont{Thiaville, Rohart,
  Émilie Jué, Cros, and Fert}}]{TRJ12}
\bibinfo{author}{\bibfnamefont{A.}~\bibnamefont{Thiaville}},
  \bibinfo{author}{\bibfnamefont{S.}~\bibnamefont{Rohart}},
  \bibinfo{author}{\bibnamefont{Émilie Jué}},
  \bibinfo{author}{\bibfnamefont{V.}~\bibnamefont{Cros}}, \bibnamefont{and}
  \bibinfo{author}{\bibfnamefont{A.}~\bibnamefont{Fert}}, \bibinfo{journal}{EPL
  (Europhysics Letters)} \textbf{\bibinfo{volume}{100}}, \bibinfo{pages}{57002}
  (\bibinfo{year}{2012}),
  \urlprefix\url{http://stacks.iop.org/0295-5075/100/i=5/a=57002}.

\bibitem[{\citenamefont{Emori et~al.}(2013)\citenamefont{Emori, Bauer, Ahn,
  Martinez, and Beach}}]{SUS13}
\bibinfo{author}{\bibfnamefont{S.}~\bibnamefont{Emori}},
  \bibinfo{author}{\bibfnamefont{U.}~\bibnamefont{Bauer}},
  \bibinfo{author}{\bibfnamefont{S.-M.} \bibnamefont{Ahn}},
  \bibinfo{author}{\bibfnamefont{E.}~\bibnamefont{Martinez}}, \bibnamefont{and}
  \bibinfo{author}{\bibfnamefont{G.~S.~D.} \bibnamefont{Beach}},
  \bibinfo{journal}{Nat Mater} \textbf{\bibinfo{volume}{12}},
  \bibinfo{pages}{611} (\bibinfo{year}{2013}),
  \urlprefix\url{http://dx.doi.org/10.1038/nmat3675}.

\bibitem[{\citenamefont{Miron et~al.}(2011)\citenamefont{Miron, Moore,
  Szambolics, Buda-Prejbeanu, Auffret, Rodmacq, Piazzini, Vogel, Bomfim, Schuhl
  et~al.}}]{ITH11}
\bibinfo{author}{\bibfnamefont{I.~M.} \bibnamefont{Miron}},
  \bibinfo{author}{\bibfnamefont{T.}~\bibnamefont{Moore}},
  \bibinfo{author}{\bibfnamefont{H.}~\bibnamefont{Szambolics}},
  \bibinfo{author}{\bibfnamefont{L.~D.} \bibnamefont{Buda-Prejbeanu}},
  \bibinfo{author}{\bibfnamefont{S.}~\bibnamefont{Auffret}},
  \bibinfo{author}{\bibfnamefont{B.}~\bibnamefont{Rodmacq}},
  \bibinfo{author}{\bibfnamefont{S.}~\bibnamefont{Piazzini}},
  \bibinfo{author}{\bibfnamefont{J.}~\bibnamefont{Vogel}},
  \bibinfo{author}{\bibfnamefont{M.}~\bibnamefont{Bomfim}},
  \bibinfo{author}{\bibfnamefont{A.}~\bibnamefont{Schuhl}},
  \bibnamefont{et~al.}, \bibinfo{journal}{Nat Mater.}
  \textbf{\bibinfo{volume}{10}}, \bibinfo{pages}{419} (\bibinfo{year}{2011}).

\bibitem[{\citenamefont{Checkelsky et~al.}(2012)\citenamefont{Checkelsky, Ye,
  Onose, Iwasa, and Tokura}}]{CYO12}
\bibinfo{author}{\bibfnamefont{J.~G.} \bibnamefont{Checkelsky}},
  \bibinfo{author}{\bibfnamefont{J.}~\bibnamefont{Ye}},
  \bibinfo{author}{\bibfnamefont{Y.}~\bibnamefont{Onose}},
  \bibinfo{author}{\bibfnamefont{Y.}~\bibnamefont{Iwasa}}, \bibnamefont{and}
  \bibinfo{author}{\bibfnamefont{Y.}~\bibnamefont{Tokura}},
  \bibinfo{journal}{Nat Phys} \textbf{\bibinfo{volume}{8}},
  \bibinfo{pages}{729} (\bibinfo{year}{2012}),
  \urlprefix\url{http://dx.doi.org/10.1038/nphys2388}.

\bibitem[{\citenamefont{Burkov and Hawthorn}(2010)}]{BH10}
\bibinfo{author}{\bibfnamefont{A.~A.} \bibnamefont{Burkov}} \bibnamefont{and}
  \bibinfo{author}{\bibfnamefont{D.~G.} \bibnamefont{Hawthorn}},
  \bibinfo{journal}{Phys. Rev. Lett.} \textbf{\bibinfo{volume}{105}},
  \bibinfo{pages}{066802} (\bibinfo{year}{2010}),
  \urlprefix\url{http://link.aps.org/doi/10.1103/PhysRevLett.105.066802}.

\bibitem[{\citenamefont{Culcer et~al.}(2010)\citenamefont{Culcer, Hwang,
  Stanescu, and Das~Sarma}}]{CHS10}
\bibinfo{author}{\bibfnamefont{D.}~\bibnamefont{Culcer}},
  \bibinfo{author}{\bibfnamefont{E.~H.} \bibnamefont{Hwang}},
  \bibinfo{author}{\bibfnamefont{T.~D.} \bibnamefont{Stanescu}},
  \bibnamefont{and}
  \bibinfo{author}{\bibfnamefont{S.}~\bibnamefont{Das~Sarma}},
  \bibinfo{journal}{Phys. Rev. B} \textbf{\bibinfo{volume}{82}},
  \bibinfo{pages}{155457} (\bibinfo{year}{2010}),
  \urlprefix\url{http://link.aps.org/doi/10.1103/PhysRevB.82.155457}.

\bibitem[{\citenamefont{Pesin and MacDonald}(2012)}]{PM12}
\bibinfo{author}{\bibfnamefont{D.}~\bibnamefont{Pesin}} \bibnamefont{and}
  \bibinfo{author}{\bibfnamefont{A.~H.} \bibnamefont{MacDonald}},
  \bibinfo{journal}{Nat Mater} \textbf{\bibinfo{volume}{11}},
  \bibinfo{pages}{409} (\bibinfo{year}{2012}),
  \urlprefix\url{http://dx.doi.org/10.1038/nmat3305}.

\bibitem[{\citenamefont{{Fischer} et~al.}(2013)\citenamefont{{Fischer},
  {Vaezi}, {Manchon}, and {Kim}}}]{FVM13}
\bibinfo{author}{\bibfnamefont{M.~H.} \bibnamefont{{Fischer}}},
  \bibinfo{author}{\bibfnamefont{A.}~\bibnamefont{{Vaezi}}},
  \bibinfo{author}{\bibfnamefont{A.}~\bibnamefont{{Manchon}}},
  \bibnamefont{and} \bibinfo{author}{\bibfnamefont{E.-A.} \bibnamefont{{Kim}}},
  \bibinfo{journal}{ArXiv e-prints}  (\bibinfo{year}{2013}),
  \eprint{1305.1328}.

\bibitem[{\citenamefont{Fan et~al.}(2014)\citenamefont{Fan, Upadhyaya, Kou,
  Lang, Takei, Wang, Tang, He, Chang, Montazeri et~al.}}]{FUK14}
\bibinfo{author}{\bibfnamefont{Y.}~\bibnamefont{Fan}},
  \bibinfo{author}{\bibfnamefont{P.}~\bibnamefont{Upadhyaya}},
  \bibinfo{author}{\bibfnamefont{X.}~\bibnamefont{Kou}},
  \bibinfo{author}{\bibfnamefont{M.}~\bibnamefont{Lang}},
  \bibinfo{author}{\bibfnamefont{S.}~\bibnamefont{Takei}},
  \bibinfo{author}{\bibfnamefont{Z.}~\bibnamefont{Wang}},
  \bibinfo{author}{\bibfnamefont{J.}~\bibnamefont{Tang}},
  \bibinfo{author}{\bibfnamefont{L.}~\bibnamefont{He}},
  \bibinfo{author}{\bibfnamefont{L.-T.} \bibnamefont{Chang}},
  \bibinfo{author}{\bibfnamefont{M.}~\bibnamefont{Montazeri}},
  \bibnamefont{et~al.}, \bibinfo{journal}{Nat Mater}
  \textbf{\bibinfo{volume}{13}}, \bibinfo{pages}{699} (\bibinfo{year}{2014}),
  \urlprefix\url{http://dx.doi.org/10.1038/nmat3973}.

\bibitem[{\citenamefont{Mellnik et~al.}(2014)\citenamefont{Mellnik, Lee,
  Richardella, Grab, Mintun, Fischer, Vaezi, Manchon, Kim, Samarth
  et~al.}}]{MLR14}
\bibinfo{author}{\bibfnamefont{A.~R.} \bibnamefont{Mellnik}},
  \bibinfo{author}{\bibfnamefont{J.~S.} \bibnamefont{Lee}},
  \bibinfo{author}{\bibfnamefont{A.}~\bibnamefont{Richardella}},
  \bibinfo{author}{\bibfnamefont{J.~L.} \bibnamefont{Grab}},
  \bibinfo{author}{\bibfnamefont{P.~J.} \bibnamefont{Mintun}},
  \bibinfo{author}{\bibfnamefont{M.~H.} \bibnamefont{Fischer}},
  \bibinfo{author}{\bibfnamefont{A.}~\bibnamefont{Vaezi}},
  \bibinfo{author}{\bibfnamefont{A.}~\bibnamefont{Manchon}},
  \bibinfo{author}{\bibfnamefont{E.~A.} \bibnamefont{Kim}},
  \bibinfo{author}{\bibfnamefont{N.}~\bibnamefont{Samarth}},
  \bibnamefont{et~al.}, \bibinfo{journal}{Nature}
  \textbf{\bibinfo{volume}{511}}, \bibinfo{pages}{449} (\bibinfo{year}{2014}),
  \urlprefix\url{http://dx.doi.org/10.1038/nature13534}.

\bibitem[{\citenamefont{Yokoyama
  et~al.}(2010{\natexlab{a}})\citenamefont{Yokoyama, Tanaka, and
  Nagaosa}}]{YTN10}
\bibinfo{author}{\bibfnamefont{T.}~\bibnamefont{Yokoyama}},
  \bibinfo{author}{\bibfnamefont{Y.}~\bibnamefont{Tanaka}}, \bibnamefont{and}
  \bibinfo{author}{\bibfnamefont{N.}~\bibnamefont{Nagaosa}},
  \bibinfo{journal}{Phys. Rev. B} \textbf{\bibinfo{volume}{81}},
  \bibinfo{pages}{121401} (\bibinfo{year}{2010}{\natexlab{a}}),
  \urlprefix\url{http://link.aps.org/doi/10.1103/PhysRevB.81.121401}.

\bibitem[{\citenamefont{Yokoyama
  et~al.}(2010{\natexlab{b}})\citenamefont{Yokoyama, Zang, and
  Nagaosa}}]{YZN10}
\bibinfo{author}{\bibfnamefont{T.}~\bibnamefont{Yokoyama}},
  \bibinfo{author}{\bibfnamefont{J.}~\bibnamefont{Zang}}, \bibnamefont{and}
  \bibinfo{author}{\bibfnamefont{N.}~\bibnamefont{Nagaosa}},
  \bibinfo{journal}{Phys. Rev. B} \textbf{\bibinfo{volume}{81}},
  \bibinfo{pages}{241410} (\bibinfo{year}{2010}{\natexlab{b}}),
  \urlprefix\url{http://link.aps.org/doi/10.1103/PhysRevB.81.241410}.

\bibitem[{\citenamefont{Garate and Franz}(2010)}]{GF10}
\bibinfo{author}{\bibfnamefont{I.}~\bibnamefont{Garate}} \bibnamefont{and}
  \bibinfo{author}{\bibfnamefont{M.}~\bibnamefont{Franz}},
  \bibinfo{journal}{Phys. Rev. Lett.} \textbf{\bibinfo{volume}{104}},
  \bibinfo{pages}{146802} (\bibinfo{year}{2010}),
  \urlprefix\url{http://link.aps.org/doi/10.1103/PhysRevLett.104.146802}.

\bibitem[{\citenamefont{Yokoyama}(2011)}]{Y11}
\bibinfo{author}{\bibfnamefont{T.}~\bibnamefont{Yokoyama}},
  \bibinfo{journal}{Phys. Rev. B} \textbf{\bibinfo{volume}{84}},
  \bibinfo{pages}{113407} (\bibinfo{year}{2011}),
  \urlprefix\url{http://link.aps.org/doi/10.1103/PhysRevB.84.113407}.

\bibitem[{\citenamefont{Ueda et~al.}(2012)\citenamefont{Ueda, Takeuchi, Tatara,
  and Yokoyama}}]{UTT12}
\bibinfo{author}{\bibfnamefont{H.~T.} \bibnamefont{Ueda}},
  \bibinfo{author}{\bibfnamefont{A.}~\bibnamefont{Takeuchi}},
  \bibinfo{author}{\bibfnamefont{G.}~\bibnamefont{Tatara}}, \bibnamefont{and}
  \bibinfo{author}{\bibfnamefont{T.}~\bibnamefont{Yokoyama}},
  \bibinfo{journal}{Phys. Rev. B} \textbf{\bibinfo{volume}{85}},
  \bibinfo{pages}{115110} (\bibinfo{year}{2012}),
  \urlprefix\url{http://link.aps.org/doi/10.1103/PhysRevB.85.115110}.

\bibitem[{\citenamefont{Nogueira and Eremin}(2012)}]{NE12}
\bibinfo{author}{\bibfnamefont{F.~S.} \bibnamefont{Nogueira}} \bibnamefont{and}
  \bibinfo{author}{\bibfnamefont{I.}~\bibnamefont{Eremin}},
  \bibinfo{journal}{Phys. Rev. Lett.} \textbf{\bibinfo{volume}{109}},
  \bibinfo{pages}{237203} (\bibinfo{year}{2012}),
  \urlprefix\url{http://link.aps.org/doi/10.1103/PhysRevLett.109.237203}.

\bibitem[{\citenamefont{Nogueira and Eremin}(2014)}]{NE14}
\bibinfo{author}{\bibfnamefont{F.~S.} \bibnamefont{Nogueira}} \bibnamefont{and}
  \bibinfo{author}{\bibfnamefont{I.}~\bibnamefont{Eremin}},
  \bibinfo{journal}{Phys. Rev. B} \textbf{\bibinfo{volume}{90}},
  \bibinfo{pages}{014431} (\bibinfo{year}{2014}),
  \urlprefix\url{http://link.aps.org/doi/10.1103/PhysRevB.90.014431}.

\bibitem[{\citenamefont{Tserkovnyak et~al.}(2015)\citenamefont{Tserkovnyak,
  Pesin, and Loss}}]{TPL15}
\bibinfo{author}{\bibfnamefont{Y.}~\bibnamefont{Tserkovnyak}},
  \bibinfo{author}{\bibfnamefont{D.~A.} \bibnamefont{Pesin}}, \bibnamefont{and}
  \bibinfo{author}{\bibfnamefont{D.}~\bibnamefont{Loss}},
  \bibinfo{journal}{Phys. Rev. B} \textbf{\bibinfo{volume}{91}},
  \bibinfo{pages}{041121} (\bibinfo{year}{2015}),
  \urlprefix\url{http://link.aps.org/doi/10.1103/PhysRevB.91.041121}.

\bibitem[{\citenamefont{Nomura and Nagaosa}(2010)}]{NN10}
\bibinfo{author}{\bibfnamefont{K.}~\bibnamefont{Nomura}} \bibnamefont{and}
  \bibinfo{author}{\bibfnamefont{N.}~\bibnamefont{Nagaosa}},
  \bibinfo{journal}{Phys. Rev. B} \textbf{\bibinfo{volume}{82}},
  \bibinfo{pages}{161401} (\bibinfo{year}{2010}),
  \urlprefix\url{http://link.aps.org/doi/10.1103/PhysRevB.82.161401}.

\bibitem[{\citenamefont{Tserkovnyak and Loss}(2012)}]{TL12}
\bibinfo{author}{\bibfnamefont{Y.}~\bibnamefont{Tserkovnyak}} \bibnamefont{and}
  \bibinfo{author}{\bibfnamefont{D.}~\bibnamefont{Loss}},
  \bibinfo{journal}{Phys. Rev. Lett.} \textbf{\bibinfo{volume}{108}},
  \bibinfo{pages}{187201} (\bibinfo{year}{2012}),
  \urlprefix\url{http://link.aps.org/doi/10.1103/PhysRevLett.108.187201}.

\bibitem[{\citenamefont{Wickles and Belzig}(2012)}]{WB12}
\bibinfo{author}{\bibfnamefont{C.}~\bibnamefont{Wickles}} \bibnamefont{and}
  \bibinfo{author}{\bibfnamefont{W.}~\bibnamefont{Belzig}},
  \bibinfo{journal}{Phys. Rev. B} \textbf{\bibinfo{volume}{86}},
  \bibinfo{pages}{035151} (\bibinfo{year}{2012}),
  \urlprefix\url{http://link.aps.org/doi/10.1103/PhysRevB.86.035151}.

\bibitem[{\citenamefont{Hammer and P\"otz}(2013)}]{HP13}
\bibinfo{author}{\bibfnamefont{R.}~\bibnamefont{Hammer}} \bibnamefont{and}
  \bibinfo{author}{\bibfnamefont{W.}~\bibnamefont{P\"otz}},
  \bibinfo{journal}{Phys. Rev. B} \textbf{\bibinfo{volume}{88}},
  \bibinfo{pages}{235119} (\bibinfo{year}{2013}),
  \urlprefix\url{http://link.aps.org/doi/10.1103/PhysRevB.88.235119}.

\bibitem[{\citenamefont{Ferreiros and Cortijo}(2014)}]{FC14}
\bibinfo{author}{\bibfnamefont{Y.}~\bibnamefont{Ferreiros}} \bibnamefont{and}
  \bibinfo{author}{\bibfnamefont{A.}~\bibnamefont{Cortijo}},
  \bibinfo{journal}{Phys. Rev. B} \textbf{\bibinfo{volume}{89}},
  \bibinfo{pages}{024413} (\bibinfo{year}{2014}),
  \urlprefix\url{http://link.aps.org/doi/10.1103/PhysRevB.89.024413}.

\bibitem[{\citenamefont{Linder}(2014)}]{L14}
\bibinfo{author}{\bibfnamefont{J.}~\bibnamefont{Linder}},
  \bibinfo{journal}{Phys. Rev. B} \textbf{\bibinfo{volume}{90}},
  \bibinfo{pages}{041412} (\bibinfo{year}{2014}),
  \urlprefix\url{http://link.aps.org/doi/10.1103/PhysRevB.90.041412}.

\bibitem[{\citenamefont{{Wakatsuki} et~al.}(2014)\citenamefont{{Wakatsuki},
  {Ezawa}, and {Nagaosa}}}]{WEN14}
\bibinfo{author}{\bibfnamefont{R.}~\bibnamefont{{Wakatsuki}}},
  \bibinfo{author}{\bibfnamefont{M.}~\bibnamefont{{Ezawa}}}, \bibnamefont{and}
  \bibinfo{author}{\bibfnamefont{N.}~\bibnamefont{{Nagaosa}}},
  \bibinfo{journal}{ArXiv e-prints}  (\bibinfo{year}{2014}),
  \eprint{1412.7910}.

\bibitem[{\citenamefont{Callan and Harvey}(1985)}]{CH85}
\bibinfo{author}{\bibfnamefont{C.~G.} \bibnamefont{Callan}} \bibnamefont{and}
  \bibinfo{author}{\bibfnamefont{J.~A.} \bibnamefont{Harvey}},
  \bibinfo{journal}{Nuclear Physics B} \textbf{\bibinfo{volume}{250}},
  \bibinfo{pages}{427 } (\bibinfo{year}{1985}).

\bibitem[{\citenamefont{Chandrasekharan}(1994)}]{C94}
\bibinfo{author}{\bibfnamefont{S.}~\bibnamefont{Chandrasekharan}},
  \bibinfo{journal}{Phys. Rev. D} \textbf{\bibinfo{volume}{49}},
  \bibinfo{pages}{1980} (\bibinfo{year}{1994}).

\bibitem[{\citenamefont{Jackiw and Rajaraman}(1985)}]{JR85}
\bibinfo{author}{\bibfnamefont{R.}~\bibnamefont{Jackiw}} \bibnamefont{and}
  \bibinfo{author}{\bibfnamefont{R.}~\bibnamefont{Rajaraman}},
  \bibinfo{journal}{Phys. Rev. Lett.} \textbf{\bibinfo{volume}{54}},
  \bibinfo{pages}{1219} (\bibinfo{year}{1985}),
  \urlprefix\url{http://link.aps.org/doi/10.1103/PhysRevLett.54.1219}.

\bibitem[{\citenamefont{Itoyama and Mueller}(1983)}]{HM83}
\bibinfo{author}{\bibfnamefont{H.}~\bibnamefont{Itoyama}} \bibnamefont{and}
  \bibinfo{author}{\bibfnamefont{A.}~\bibnamefont{Mueller}},
  \bibinfo{journal}{Nuclear Physics B} \textbf{\bibinfo{volume}{218}},
  \bibinfo{pages}{349 } (\bibinfo{year}{1983}), ISSN \bibinfo{issn}{0550-3213},
  \urlprefix\url{http://www.sciencedirect.com/science/article/pii/055032138390370X}.

\bibitem[{\citenamefont{Das and Karev}(1987)}]{DK87}
\bibinfo{author}{\bibfnamefont{A.}~\bibnamefont{Das}} \bibnamefont{and}
  \bibinfo{author}{\bibfnamefont{A.}~\bibnamefont{Karev}},
  \bibinfo{journal}{Phys. Rev. D} \textbf{\bibinfo{volume}{36}},
  \bibinfo{pages}{623} (\bibinfo{year}{1987}),
  \urlprefix\url{http://link.aps.org/doi/10.1103/PhysRevD.36.623}.

\bibitem[{\citenamefont{Redlich}(1984)}]{R84}
\bibinfo{author}{\bibfnamefont{A.~N.} \bibnamefont{Redlich}},
  \bibinfo{journal}{Phys. Rev. D} \textbf{\bibinfo{volume}{29}},
  \bibinfo{pages}{2366} (\bibinfo{year}{1984}),
  \urlprefix\url{http://link.aps.org/doi/10.1103/PhysRevD.29.2366}.

\bibitem[{\citenamefont{Brandt et~al.}(2000)\citenamefont{Brandt, Das, and
  Frenkel}}]{BDF00}
\bibinfo{author}{\bibfnamefont{F.~T.} \bibnamefont{Brandt}},
  \bibinfo{author}{\bibfnamefont{A.}~\bibnamefont{Das}}, \bibnamefont{and}
  \bibinfo{author}{\bibfnamefont{J.}~\bibnamefont{Frenkel}},
  \bibinfo{journal}{Phys. Rev. D} \textbf{\bibinfo{volume}{62}},
  \bibinfo{pages}{085012} (\bibinfo{year}{2000}),
  \urlprefix\url{http://link.aps.org/doi/10.1103/PhysRevD.62.085012}.

\bibitem[{\citenamefont{Dunne et~al.}(1997)\citenamefont{Dunne, Lee, and
  Lu}}]{DLL97}
\bibinfo{author}{\bibfnamefont{G.}~\bibnamefont{Dunne}},
  \bibinfo{author}{\bibfnamefont{K.}~\bibnamefont{Lee}}, \bibnamefont{and}
  \bibinfo{author}{\bibfnamefont{C.}~\bibnamefont{Lu}}, \bibinfo{journal}{Phys.
  Rev. Lett.} \textbf{\bibinfo{volume}{78}}, \bibinfo{pages}{3434}
  (\bibinfo{year}{1997}),
  \urlprefix\url{http://link.aps.org/doi/10.1103/PhysRevLett.78.3434}.

\bibitem[{\citenamefont{Deser et~al.}(1997)\citenamefont{Deser, Griguolo, and
  Seminara}}]{DGS97}
\bibinfo{author}{\bibfnamefont{S.}~\bibnamefont{Deser}},
  \bibinfo{author}{\bibfnamefont{L.}~\bibnamefont{Griguolo}}, \bibnamefont{and}
  \bibinfo{author}{\bibfnamefont{D.}~\bibnamefont{Seminara}},
  \bibinfo{journal}{Phys. Rev. Lett.} \textbf{\bibinfo{volume}{79}},
  \bibinfo{pages}{1976} (\bibinfo{year}{1997}),
  \urlprefix\url{http://link.aps.org/doi/10.1103/PhysRevLett.79.1976}.

\bibitem[{\citenamefont{Brandt et~al.}(2001)\citenamefont{Brandt, Das, Frenkel,
  Pereira, and Taylor}}]{BDF01}
\bibinfo{author}{\bibfnamefont{F.~T.} \bibnamefont{Brandt}},
  \bibinfo{author}{\bibfnamefont{A.}~\bibnamefont{Das}},
  \bibinfo{author}{\bibfnamefont{J.}~\bibnamefont{Frenkel}},
  \bibinfo{author}{\bibfnamefont{S.}~\bibnamefont{Pereira}}, \bibnamefont{and}
  \bibinfo{author}{\bibfnamefont{J.~C.} \bibnamefont{Taylor}},
  \bibinfo{journal}{Phys. Rev. D} \textbf{\bibinfo{volume}{64}},
  \bibinfo{pages}{065018} (\bibinfo{year}{2001}),
  \urlprefix\url{http://link.aps.org/doi/10.1103/PhysRevD.64.065018}.

\bibitem[{\citenamefont{Tatara et~al.}(2008)\citenamefont{Tatara, Kohno, and
  Shibata}}]{TKS08}
\bibinfo{author}{\bibfnamefont{G.}~\bibnamefont{Tatara}},
  \bibinfo{author}{\bibfnamefont{H.}~\bibnamefont{Kohno}}, \bibnamefont{and}
  \bibinfo{author}{\bibfnamefont{J.}~\bibnamefont{Shibata}},
  \bibinfo{journal}{Physics Reports} \textbf{\bibinfo{volume}{468}},
  \bibinfo{pages}{213 } (\bibinfo{year}{2008}), ISSN \bibinfo{issn}{0370-1573},
  \urlprefix\url{http://www.sciencedirect.com/science/article/pii/S0370157308002597}.

\bibitem[{\citenamefont{Shibata et~al.}(2011)\citenamefont{Shibata, Tatara, and
  Kohno}}]{STK11}
\bibinfo{author}{\bibfnamefont{J.}~\bibnamefont{Shibata}},
  \bibinfo{author}{\bibfnamefont{G.}~\bibnamefont{Tatara}}, \bibnamefont{and}
  \bibinfo{author}{\bibfnamefont{H.}~\bibnamefont{Kohno}},
  \bibinfo{journal}{Journal of Physics D: Applied Physics}
  \textbf{\bibinfo{volume}{44}}, \bibinfo{pages}{384004}
  (\bibinfo{year}{2011}),
  \urlprefix\url{http://stacks.iop.org/0022-3727/44/i=38/a=384004}.

\bibitem[{\citenamefont{Martinez et~al.}(2009)\citenamefont{Martinez,
  Lopez-Diaz, Alejos, and Torres}}]{MLA09}
\bibinfo{author}{\bibfnamefont{E.}~\bibnamefont{Martinez}},
  \bibinfo{author}{\bibfnamefont{L.}~\bibnamefont{Lopez-Diaz}},
  \bibinfo{author}{\bibfnamefont{O.}~\bibnamefont{Alejos}}, \bibnamefont{and}
  \bibinfo{author}{\bibfnamefont{L.}~\bibnamefont{Torres}},
  \bibinfo{journal}{Journal of Applied Physics} \textbf{\bibinfo{volume}{106}},
  \bibinfo{eid}{043914} (\bibinfo{year}{2009}),
  \urlprefix\url{http://scitation.aip.org/content/aip/journal/jap/106/4/10.1063/1.3204496}.

\bibitem[{\citenamefont{Xiu et~al.}(2011)\citenamefont{Xiu, He, Wang, Cheng,
  Chang, Lang, Huang, Kou, Zhou, Jiang et~al.}}]{XHW11}
\bibinfo{author}{\bibfnamefont{F.}~\bibnamefont{Xiu}},
  \bibinfo{author}{\bibfnamefont{L.}~\bibnamefont{He}},
  \bibinfo{author}{\bibfnamefont{Y.}~\bibnamefont{Wang}},
  \bibinfo{author}{\bibfnamefont{L.}~\bibnamefont{Cheng}},
  \bibinfo{author}{\bibfnamefont{L.-T.} \bibnamefont{Chang}},
  \bibinfo{author}{\bibfnamefont{M.}~\bibnamefont{Lang}},
  \bibinfo{author}{\bibfnamefont{G.}~\bibnamefont{Huang}},
  \bibinfo{author}{\bibfnamefont{X.}~\bibnamefont{Kou}},
  \bibinfo{author}{\bibfnamefont{Y.}~\bibnamefont{Zhou}},
  \bibinfo{author}{\bibfnamefont{X.}~\bibnamefont{Jiang}},
  \bibnamefont{et~al.}, \bibinfo{journal}{Nat Nano}
  \textbf{\bibinfo{volume}{6}}, \bibinfo{pages}{216} (\bibinfo{year}{2011}),
  \urlprefix\url{http://dx.doi.org/10.1038/nnano.2011.19}.

\bibitem[{\citenamefont{Fl\"{u}gge}(1998)}]{F98}
\bibinfo{author}{\bibfnamefont{S.}~\bibnamefont{Fl\"{u}gge}},
  \bibinfo{journal}{Springer}  (\bibinfo{year}{1998}).

\bibitem[{\citenamefont{Kapusta and Gale}(2006)}]{KG06}
\bibinfo{author}{\bibfnamefont{J.~I.} \bibnamefont{Kapusta}} \bibnamefont{and}
  \bibinfo{author}{\bibfnamefont{C.}~\bibnamefont{Gale}},
  \bibinfo{journal}{Cambridge University Press}  (\bibinfo{year}{2006}).

\bibitem[{\citenamefont{Scholz and Schliemann}(2011)}]{SS11}
\bibinfo{author}{\bibfnamefont{A.}~\bibnamefont{Scholz}} \bibnamefont{and}
  \bibinfo{author}{\bibfnamefont{J.}~\bibnamefont{Schliemann}},
  \bibinfo{journal}{Phys. Rev. B} \textbf{\bibinfo{volume}{83}},
  \bibinfo{pages}{235409} (\bibinfo{year}{2011}),
  \urlprefix\url{http://link.aps.org/doi/10.1103/PhysRevB.83.235409}.

\end{thebibliography}

\end{document}